%% file: ms.tex
\pdfoutput=1
\documentclass{vldb}
\usepackage{ifpdf}
\usepackage{graphicx}
\usepackage{balance}  
\usepackage{cite}
\usepackage[linesnumbered,ruled,lined]{algorithm2e}
\usepackage{amsmath,amssymb,amsfonts}
\usepackage{algorithmic}
\usepackage{graphicx}
\usepackage{textcomp}
\usepackage{xcolor}
\usepackage{xspace,colortbl,subfigure,multirow}
\usepackage{bm}
\usepackage{url}

\newcommand{\eat}[1]{\xspace}
\newcommand{\myarch}{Poplar\xspace}
\newcommand{\lsn}{LSN\xspace}
\newcommand{\gsn}{GSN\xspace}
\newcommand{\mylsn}{SSN\xspace}
\newcommand{\mylsns}{SSNs\xspace}
\newcommand{\centralized}{{\small \textsf{CENTR}}\xspace}
\newcommand{\nvmd}{{\small \textsf{NVM-D}}\xspace}
\newcommand{\para}{{\small \textsf{POPLAR}}\xspace}
\newcommand{\silo}{{\small \textsf{SILO}}\xspace}
\newdef{definition}{\textsc{Level}}

\vldbTitle{}
\vldbAuthors{}
\vldbDOI{}
\vldbVolume{}
\vldbNumber{}
\vldbYear{}

\begin{document}

\title{Guaranteeing Recoverability via Partially Constrained Transaction Logs}




%
%
%
%

\numberofauthors{1} 

\author{
%
%
\alignauthor
Huan Zhou, Jinwei Guo, Huiqi Hu, Weining Qian, Xuan Zhou, Aoying Zhou\\
       \affaddr{School of Data Science \& Engineering}\\
       \affaddr{East China Normal University}\\
       \email{\{zhouhuan,guojinwei\}@stu.ecnu.edu.cn, \{hqhu,wnqian,xzhou,ayzhou\}@dase.ecnu.edu.cn}
}

\maketitle
\input{abstract.tex}
\input{introduction.tex}

\input{background.tex}

\input{recoverability.tex}

\input{logging.tex}

\input{recovery.tex}

\input{evaluation.tex}

\input{relatedwork.tex}

\input{conclusion.tex}

\balance

\bibliographystyle{habbrv}
\bibliography{ms}  


\end{document}

%% file: abstract.tex
\begin{abstract}
Transaction logging is an essential constituent to guarantee the atomicity and 
durability in online transaction processing (OLTP) systems. 
It always has a considerable impact on performance, especially in
an in-memory database system. 
Conventional implementations of logging rely heavily on a centralized design, 
which guarantees the 
correctness of recovery by enforcing a total order of all operations such as log sequence 
number (\lsn) allocation, log persistence, transaction committing and 
recovering. This strict sequential constraint seriously 
limits the scalability and parallelism of transaction logging and recovery, 
especially in the multi-core hardware environment. 

In this paper, we define recoverability for transaction logging and demonstrate its correctness for crash recovery. 
Based on recoverability, we propose a recoverable logging scheme named 
\myarch,  
which enables scalable and parallel log processing by easing the restrictions. 
Its main advantages are that 
(1) \myarch enables the parallel log persistence on multiple storage devices; 
(2) it replaces the centralized \lsn allocation by calculating a partially 
ordered sequence number in a distributed manner, which allows log records to 
only track RAW and WAW dependencies among transactions; 
(3) it only demands transactions with RAW dependencies to be committed in serial order; 
(4) \myarch can concurrently restore a consistent database state based on the partially constrained logs after a crash. 
Experimental results show that \myarch scales well with the increase of IO 
devices and outperforms other logging approaches on both SSDs and emulated non-volatile memory. 
\end{abstract}

%% file: introduction.tex
\section{Introduction}
Transaction logging as an established recovery method is widely used in 
database management systems (DBMSs) to ensure atomicity and durability of 
transactions. 
In in-memory DBMSs, it persists information of committed transactions in the form of log records 
and recovers them in a proper order to restore a consistent database state after a system crash. 
As the only component that involves persistence and dependency tracking, 
transaction logging often dominates the system throughput and latency when executing 
highly concurrent online transaction processing (OLTP) workloads. 
Most in-memory DBMSs implement a variant of ARIES-Logging~\cite{VLDB:mohan1992aries}, 
which first caches log records in a central, volatile log buffer and then 
forces them into a single storage device before transactions commit.  
Each log record is assigned with a unique and monotonically increasing log 
sequence number (LSN) that determines the order of persisting log records, the 
order of committing transactions and the order of recovering transactions.

The centralized design of transaction logging becomes the main performance 
bottleneck in multi-core platforms  
and new storage devices, such as non-volatile memory (NVM). 
There are three key obstacles:  
(1) \lsn allocation based on the centralized log buffer is a notorious contention point that
seriously hinders scalability; 
(2) flushing log records into a single storage device limits maximal performance due 
to the limited IO bandwidth, especially with a slower storage device; 
(3) there is a sequential constraint in centralized logging that demands \lsn allocation, log persistence, transaction committing and recovering follow a same total order, which terribly limits the parallelism of logging and recovery. 
Prior works~\cite{VLDB:johnson2010aether,VLDB:jung2017scalable} have addressed 
the first bottleneck with notable improvements, 
but the global contention still exists due to the central log buffer used in those
proposed solutions. 
Other studies~\cite{VLDB:wang2014scalable,VLDB:zheng2014fast}
use multiple log buffers instead of the central log buffer and allow log 
records to be concurrently written into multiple storage devices, which 
eliminates centralized contention and  IO bandwidth limitation. 
But these solutions either allow worker threads to force log records directly to storage devices, or use a coarse-grained commit protocol, 
which is not suitable for slower storage devices (e.g., 
SSD, HDD) and extends the commit latency. 

So far, all previous researches have not addressed the third issue.  
Operations such as \lsn allocation and transaction committing in their 
approaches are serial or at least follow all dependencies among transactions. 
But in fact, the strict sequential constraint for crash recovery 
is not necessary, especially for transactions without any dependencies. 
In this paper, we analyze comprehensively which constraints on transaction logging are essential for correctness during recovery, 
and gives a definition of \emph{recoverability} for transaction logging. 
The recoverability requires that the commit order of transactions tracks the read-after-write (RAW) dependencies among transactions and the sequence number of log records complies with the write-after-write (WAW) dependencies among transactions. 

Based on the definition of recoverability, we propose a scalable transaction logging, named 
as \myarch, that can perform both logging and recovery in parallel. 
The centerpiece of \myarch is to remove the sequential constraint, 
which allows 
(1) log records to be written into multiple storage devices in non-serial order; 
(2) only transactions with RAW dependencies to be committed in serial order; 
(3) sequence number of log records to track only WAW dependencies among transactions. 
To track these dependencies without additional overheads, 
we use a scalable sequence number (\mylsn) instead of centralized \lsn to track the WAW and RAW dependencies.  
The \mylsn is calculated in a decentralized manner, which exhibits high scalability. 
Based on \mylsn , \myarch can identify directly whether transactions with RAW dependencies can be committed. 
After a system crash, \myarch can recover persistent log records in \mylsn order to restore a consistent database state. 
In addition, combined with optimistic concurrency control, \mylsn can be used as the commit timestamp of transactions to guarantee serializability, while removing the centralized contention on the timestamp allocation. 

We implement \myarch in the open-source database system DBx1000~\cite{VLDB:DBx1000} and 
compare it with other transaction logging approaches. 
Experimental results show that both in YCSB and TPC-C benchmarks, \myarch is well scaled with the increase of IO devices and has the highest throughput with excellent commit latency of transactions. 
Moreover, our logging is generally applicable to all kinds of storage devices, such as SSD and NVM.

In summary, we make the following contributions: 

\begin{itemize}
	\item Based on the analysis of syntactical restrictions for correct 
	recovery, we define three constraint levels for transaction logging 
	and elaborate on recoverability in main-memory database systems. 
	
	\item We categorize state-of-the-art transaction logging approaches 
	according to these three levels and summarize the performance 
	bottlenecks of centralized logging and popular parallel logging methods.  
	
	\item We propose a scalable and parallel transaction logging that eliminates the sequential constraint of traditional logging, while guaranteeing that database systems can be recovered to a consistent state after a crash. 
	
	\item Experimental results show that \myarch with only $2$ pieces of SSDs
	performs up to $2\times$ higher throughput than centralized logging, 
	$\sim 280\times$ 
	higher throughput than NVM-based logging and $\sim 6\times$ shorter commit 
	latency of transactions than Silo. 
\end{itemize}

The rest of this paper is organized as follows: Section~\ref{sec:back} reviews the traditional logging and recovery and confirms the performance bottlenecks. 
Section~\ref{sec:recoverability} gives a definition of recoverability and compares our logging with other approaches. 
Section~\ref{sec:paralogging} presents the design of our parallel logging in detail,
and the implementation of parallel recovery is shown in Section~\ref{sec:recovery}. 
Section~\ref{sec:expr} describes the experimental evaluation and Section~\ref{sec:relatedwork} reviews the related work. 
Conclusion is shown in Section~\ref{sec:conclusion}.

%% file: background.tex
\section{Background}\label{sec:back}


To successfully survive from a system failure, a recovery manager should make a database  behave as if it contains all of the effects of committed transactions (redo recovery) and none of the effects of uncommitted transactions (undo recovery). 
Typically, database management systems (DBMSs) adopt transaction logs to store redo and undo information 
of transactions in the non-volatile storage. 
In an in-memory database system, undo log is not necessary, because the system does not need to flush uncommitted data into the durable storage. 
Thus, the recovery manager does not need to erase any uncommitted writes. A log file is composed of a 
sequence of log entries, each one is called a log record that is assigned with 
a unique and monotonically increasing \emph{sequence 
number}. 
Therefore, the sequence numbers can be used to order two different updates on the same tuple. 
In a nutshell, upon the system failures, a recovery manager can replay all redo log 
records in the sequence order to reconstruct the database state from the latest 
checkpoint.


Essentially, the goal of a recovery manager is to ensure 
that the recovered state preserves the effects of all committed updates 
in a serial order of the execution of transactions. It needs to ensure all the 
committed updates have been persisted before a crash and the sequence order of 
log records matches the original execution order of transactions. 
Any violation will definitely risk the correctness of recovery. 
To comply with these requirements, a variant of ARIES-Logging~\cite{VLDB:mohan1992aries} in  
an in-memory database system is implemented as follows: 
(1) it calculates the sequence number (denoted as \lsn) of each log record by 
using a global lock; (2) it caches log records in a centralized log buffer in total sequence order, and then forces them into a permanent storage device in batch~\cite{VLDB:hagmann1987reimplementing}; 
(3) a transaction can be committed until its log records and all the log 
records preceding the last log entry of the transaction have been persistent. The logging approach has three main bottlenecks: 

1) \textbf{Contention on the centralized buffer.} The \lsn allocation by a 
global lock on the centralized log buffer is a dazzling serialization 
bottleneck. Because transaction processing---which requests the 
centralized data structure concurrently---is highly parallelized, 
contentions on the central log buffer waste a lot of CPU cycles and significantly limit the scalability.

2) \textbf{Bandwidth limitation.} 
All log records are required to be sequentially written to a storage device before their transactions commit. 
Even  the sequential write  makes full use of IO properties, 
the system throughput is bounded by the IO bandwidth of 
the device, especially when transaction produces large-size log records. 
Therefore, the database logging with a single IO device 
may be less performant due to its limited IO bandwidth. 

3) \textbf{Sequential constraint.} 
A strong constraint imposed on the transaction logging is its \emph{sequentiality}~\cite{VLDB:jung2017scalable}, which strictly requires that the order of log records, log flushing and transaction committing must be the same. 
In other words, all log records are totally-ordered and persisted with the 
global \lsn, and transactions are also committed according to the \lsn order. 
This somehow forces operations such as LSN allocation, log persistence and 
transaction committing to be serial in the logging process. Furthermore, during 
crash recovery, 
restoring database state in total \lsn order also limits the level of parallelism of recovery. 

Prior works~\cite{VLDB:johnson2010aether,VLDB:jung2017scalable} address the 
contention on centralized log buffer by using a lightweight atomic instruction 
and copying transaction's logs to the log buffer in parallel. 
But other bottlenecks still exist in these approaches due to its centralized design. 
In addition, H-store~\cite{VLDB:kallman2008h} uses command 
logging instead of value logging to reduce pressure on the IO device.  
While it significantly slows down the recovery processing, as the database 
system must redo all transactions to restore a consistent state after a crash. 
 
To eliminate all above bottlenecks, a simple intuition is to adopt a parallel 
logging mechanism that forces log records over multiple stable storage devices. 
Historically, parallel logging has been prohibitive for a single node system 
due to the sequential constraint.  
In fact, a correct recovery manager does not need to guarantee the  strong 
sequentiality, especially for transactions without any dependencies. 
In the next section, we mainly discuss the necessary restrictions on parallel 
logging for a recoverable database system. In this paper, we focus on the 
recoverability of in-memory database systems that use value logging. 
Without loss of generality, we 
assume that each transaction only produces 
a single log record containing all writes of the transaction. 


%% file: recoverability.tex
\newcommand{\txn}{\ensuremath{\textit{T}}\xspace}
\newcommand{\commit}{\ensuremath{\textit{C}}\xspace}
\newcommand{\logs}{\ensuremath{\textit{L}}\xspace}
\newcommand{\readx}{\ensuremath{\textit{R}}\xspace}
\newcommand{\writex}{\ensuremath{\textit{W}}\xspace}
\newcommand{\reco}{\ensuremath{\textrm{recoverability}}\xspace}
\newcommand{\rigo}{\ensuremath{\textrm{rigorousness}}\xspace}
\newcommand{\seri}{\ensuremath{\textrm{sequentiality}}\xspace}
\newcommand{\flush}{\ensuremath{\textit{Flush}}\xspace}

\section{Recoverability}\label{sec:recoverability}

\subsection{Constraint Levels for Logging}\label{subsec:difinition}

To correctly recover from a system crash,  a  recovery manager of in-memory DBMSs should (1) identify which transactions have been committed before the crash, 
and (2) recover log records of committed transactions in a proper order to a 
most recent and consistent database state.  
The first point indicates that  
a transaction is considered to be committed only if its log record has been 
durable and 
transactions it depends on are committed. 
In other words, a transaction commits only after those transactions it depends 
on. 
For the second point, the relative order among log records of 
conflicting transactions must comply with their execution order. It indicates 
additional dependency tracking in the logging process. 
Based on different syntactical restrictions on the order of committing 
transactions and log records, we define three levels of constraints for the crash recovery 
manager. 
For ease of explanation, we use $C_i$ to present the commit operation of 
the transaction $T_i$ and use $L_i$ to denote the sequence number of this
transaction's log record. Let $C_i \prec C_j$ present that $C_i$ happens before 
$C_j$ in the sequence of commit operations. Next, we describe these three 
levels as follows. 


\begin{definition}\label{def1} (\textsc{Recoverability}). 
The manager is recoverable, for any two transactions $T_i$, $T_j$ ($i 
\neq j$): if $\txn_j$ writes on a tuple $x$ and  $\txn_i$ reads $T_j$'s update 
on the tuple $x$, then 
$\commit_j \prec \commit_i$;
and if $\txn_i$ overwrites $\txn_j$'s update, then $\logs_j < \logs_i$. 
\end{definition}

\begin{definition}\label{def2} (\textsc{Rigorousness}). 
The manager is rigorous, for any two transactions $\txn_i$, $\txn_j$ ($i \neq 
j$), 
there are conflicts between them: if $\txn_i$ depends on $\txn_j$, 
then  $\commit_j \prec \commit_i$ and $\logs_j < \logs_i$.

\end{definition}

\begin{definition}\label{def3} (\textsc{Sequentiality}). 
The manager is sequential if 
it is rigorous and additionally satisfies the following condition:  
for any two transactions  $\txn_i$, $\txn_j$ without any conflict ($i \neq j$): 
$\commit_j \prec \commit_i$ and $\logs_j < \logs_i$, or $\commit_i \prec \commit_j$ and $\logs_i < \logs_j$. 
\end{definition}

In summary, \reco requires the commit order of transactions to track read-after-write (RAW) 
dependencies among transactions, 
and the sequence number of log records to track write-after-write (WAW) dependencies. 
Rigorousness demands that both the commit order and sequence number should guarantee RAW, WAW and write-after-read (WAR) dependencies 
among transactions. 
And sequentiality claims that the commit order of all transactions and sequence 
number of all log records must be in total order. 
Although the implementation of recovery manager is simplified by forcing 
more restrictions on the order of committing transactions and log records, 
these restrictions limit the potential capability of parallelism in logging 
process.


Note that it is worthwhile to discuss the order of committing transactions in 
logging process. 
In-memory DBMSs that use early lock release (ELR) allow \emph{a pre-committed 
transaction}{\footnote{A pre-committed transaction can not be committed only if 
a crash happens, never because of transaction 
failure.}}~\cite{VLDB:dewitt1984implementation} releases its locks before it 
commits. 
This way leads to a result that  other transactions observe the pre-committed 
transaction's dirty tuples. 
To avoid an inconsistent database state after a crash, the recovery manager must ensure 
that a pre-committed transaction commits before transactions depend on it. 


\subsection{Illustrative Examples}
\begin{figure}[!t]	
	\centerline{\includegraphics[scale=.80]{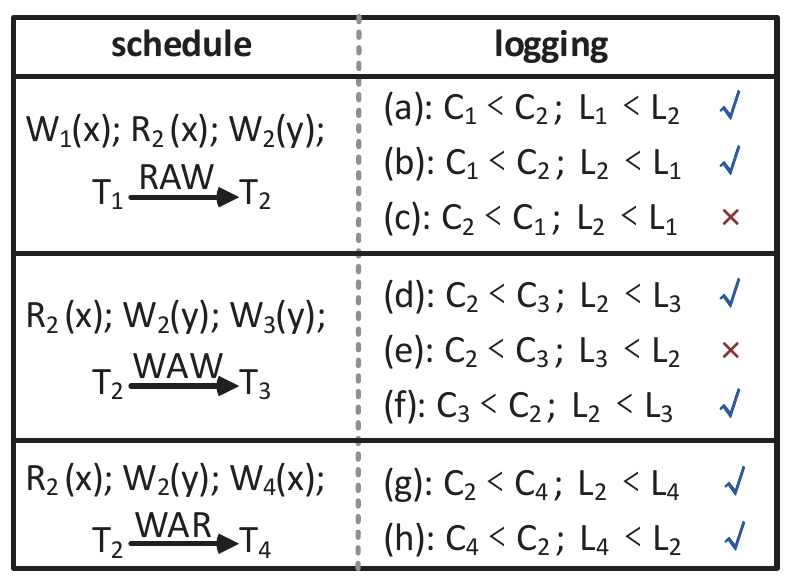}}\vspace{-1mm}
	\caption{Execution scenarios of four transaction in logging process. For the correct recovery, the order of transaction committing must track RAW dependencies and the order of log records must track WAW dependencies.}
	\label{fig:example}\vspace{-3mm}
\end{figure}

\begin{table*}[!htpb]
	\centering
	\caption{A comparison of state-of-the-art transaction logging approaches}\label{tab:1}
	\begin{tabular}{|p{1.8cm}|p{1.3cm}|p{2.5cm}|p{1.5cm}|p{1.5cm}|p{1.8cm}|p{1.8cm}|p{1.8cm}|} \hline
		&Log type&Log record&Log insert&Log flush&Txn commit&Recovery&Level\\ \hline
		ARIES~\cite{VLDB:mohan1992aries} & value & totally-ordered& serial&serial& serial& serial&  \seri\\ \hline
		Aether~\cite{VLDB:johnson2010aether} \par{\textsc{Eleda}}~\cite{VLDB:jung2017scalable} & value& totally-ordered&parallel&serial&serial& serial& \seri \\ \hline
		H-Store~\cite{VLDB:kallman2008h}& command& totally-ordered&parallel&serial& serial&serial,\par{re-execute}&\seri \\ \hline
		Silo~\cite{VLDB:tu2013speedy}& value&epoch-based totally-ordered&parallel& parallel&epoch-based serial &parallel&epoch-based \seri\\ \hline
		NV-Log~\cite{VLDB:huang2014nvram}& value&totally-ordered&N/A&parallel& serial& parallel&\seri\\ \hline
		NVM-D~\cite{VLDB:wang2014scalable}& value&RAW,WAW,\par{WAR}&N/A& parallel&RAW,WAW,\par{WAR}& parallel& \rigo\\ \hline
		\myarch& value&RAW,WAW&parallel&parallel&RAW& parallel&\reco\\ \hline
	\end{tabular}
\end{table*}

In this section, we demonstrate how \reco guarantees the correctness of databases. 

Figure~\ref{fig:example} shows possible execution scenarios of four transaction in logging process. 
The dependencies between different transactions are shown as arrows. 
In the figure, 
all transactions are pre-committed transactions that write their log records in log buffers 
but none is persistent. 
A correct symbol indicates 
that under the scenario, database system can be reconstructed to 
a consistent state in the presence of a crash, 
and the meaning of an error symbol is the opposite. 

For $\txn_1$ and $\txn_2$\ with RAW dependency, 
if we ignore the constraint on the commit order of transactions, 
$\txn_2$ can be committed before $\txn_1$, as shown in scenario(c). 
Then if the system crashes at the same time, 
only $\txn_2$ will be recovered but $\txn_1$ will not due to $\txn_1$'s log record is not durable. 
This leads to an incorrect state as $\txn_2$ observes a value of tuple $x$ that does not exist in the reconstructed database system. 
But if only the sequence number of their log records does not track the RAW dependency, 
the recovered state is alway consistent. 
As shown in scenario(b), $\txn_1$ commits before $\txn_2$ and $\txn_1$'s sequence number is larger than that of $\txn_2$. 
If a crash happens after $\txn_1$ commits but $\txn_2$ is not persistent, 
it does not result in an inconsistent state. 
And if a crash happens after two transactions are committed, 
$\txn_2$ might be recovered before $\txn_1$ due to its smaller sequence number. 
The restored state is still consistent because a read-only operation does not have any side-effect on tuples during recovery. 
Therefore, for $\txn_1$ and $\txn_2$, only $\commit_1 \prec \commit_2$ must be guarantee. 

For $\txn_2$ and $\txn_3$ with WAW dependency, 
$\txn_2$ and $\txn_3$ can be committed after their own log record has been persistent, 
but sequence number of their log records must follow the WAW dependency. 
That is to say, when $\txn_3$ overwrites $\txn_2$'s update, only $\logs_2 < \logs_3$ must be ensured. 
 If a crash happens between commit operations of two transactions in scenario(d),(e) and (f), 
only the committed transaction ($\txn_2$ or $\txn_3$) will be recovered but the other is not. 
The reconstructed state is consistent as if the uncommitted transaction has never existed. 
But, if the system crashes after all transactions have been committed, 
$\txn_2$ and $\txn_3$ are recovered in the sequence number order of their log records. 
In scenario(e), 
$\txn_2$ will be recovered after $\txn_3$, 
which leads to an inconsistent state since $\txn_2$ overwrites the value of tuple $y$ previously written by $\txn_3$ so that $\txn_3$'s update is lost. 
If their log records tracks the WAW dependency as scenario(d) and (f), 
the recovery manager can avoid the inconsistence.. 

Unlike RAW and WAW dependency, the order of committing transactions and log records do not need to track the WAR dependency. 
For $\txn_2$ and $\txn_4$, $\txn_4$ can be committed before $\txn_2$ and its sequence number can be smaller than that of $\txn_2$. 
As shown in scenario(h), 
if the system crashes after $\txn_4$ is committed but $\txn_2$ is not, 
only $\txn_4$ will be recovered during recovery. 
The database system can be reconstructed to a consistent state as if $\txn_2$ has never existed. 
And if the crash happens after two transactions have been committed,  $\txn_2$ might be recovered after $\txn_4$ due to sequence number order of their log records. 
The recovered database state is still consistent as a read does not have side-effects in tuples 
and does not affect the following write on the same tuples. 

Overall, in order to recover to a consistent database state after a crash, 
the crash recovery manager only needs to ensure that 
the commit order of transactions tracks the RAW dependencies among transactions 
and the sequence number of log records tracks the WAW dependencies among transactions. 
Therefore, the \reco level is able to guarantee system correctness after a crash.

\subsection{Comparison of Existing Approaches}


Recoverability breaks thoroughly the sequential constraint of centralized logging, 
which makes it possible to implement parallel logging in a single node system. 
Based on the recoverability level, we propose a scalable and parallel transaction logging (named as \myarch). 
So far, there are some other parallel logging approaches~\cite{VLDB:tu2013speedy,VLDB:zheng2014fast,VLDB:wang2014scalable, VLDB:huang2014nvram} based on different recovery levels. 
Table~\ref{tab:1} provides a summary comparing our method with 
various logging approaches.

\begin{figure*}[!htpb]
	\centerline{\includegraphics[scale=.80]{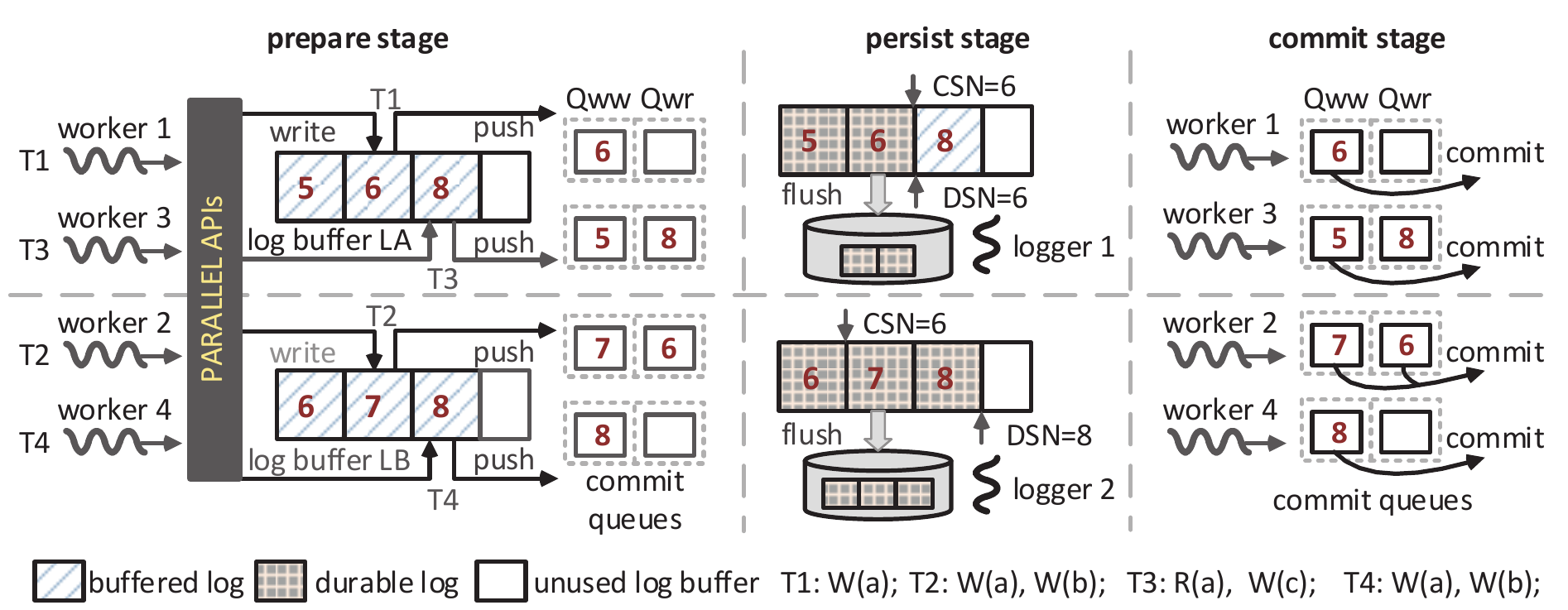}}\vspace{-2mm}
	\caption{Overview of parallel and scalable logging .}
	\label{fig:loggingoverview}\vspace{-3mm}
\end{figure*}

Silo~\cite{VLDB:tu2013speedy,VLDB:zheng2014fast} proposes a parallel transaction logging that belongs to 
sequentiality as defined in Section~\ref{subsec:difinition}.
It uses multiple log buffers to cache transactions' log records, 
and enables them to be concurrently written into multiple storage devices. 
To avoid all centralized contentions of in-memory databases, 
Silo utilizes coarse-grained epochs instead of centralized \lsn to track sequentiality among transactions,   
which exhibits excellent performance and high scalability. 
However, Silo only ensures transactions across epoch boundaries are in a serial 
order, while transactions in a same epoch can be out of order. 
To guarantee the correct recovery, it adopts an epoch-based group commit that 
persists and commits transactions in epoch units. 
Unfortunately, this way significantly increases the commit latency of transactions. 

Wang et al.~\cite{VLDB:wang2014scalable} present a distributed logging based 
on the non-volatile memory (NVM), which is called NVM-D in our paper. 
It establishes multiple log buffers on NVM and allows each worker thread to persist transactions' log records directly with the \emph{mfence} instruction. 
To ensure correctness, 
NVM-D uses a global sequence number (\gsn) to track all dependencies between transactions. 
Both the sequence number order of log records and commit order of transactions are in the \gsn order, 
which provides a rigorous recovery manager. 
The \gsn of a transaction is calculated in a distributed manner based on tuples accessed by the transaction and a log buffer stores the transaction's log record. 
NVM-D provides near-linear scalability in the NVM, 
but it is not suitable for slower storage devices because frequent \emph{mfence} will seriously hurt performance. 

NV-Logging~\cite{VLDB:huang2014nvram} is a decentralized logging, 
which allocates each worker thread a private log buffer on NVM. 
Although it allows worker threads to concurrently persist log records, 
NV-Logging also use a total \lsn to track the order of log records 
and the commit order of transactions. 
After a system crash, log records are recovered in total \lsn order to restore a consistent database state.  
Hence, it is also a sequential recovery manager.

\textbf{Design principle.} 
To address bottleneck issues of centralized transaction logging, 
we need to design a scalable and parallel transaction logging, 
which belongs to the recoverability level. 
In this paper, the new logging approach is referred to as \myarch, whose design 
principles are as follows. 
It uses multiple log buffers to first cache log records in memory 
and then forces concurrently them to multiple storage devices in batch. 
For the correctness of recovery, \myarch uses a scalable sequence number (\mylsn) to track the WAW dependencies among transactions 
and allows the commit order of transactions to follow only RAW dependencies among transactions. 
To avoid additional overhead of transaction committing, 
\mylsn also tracks RAW dependencies. 
To further address the issue of centralization bottleneck, the \mylsn is 
calculated in a decentralized 
manner, which makes \myarch highly scalable. 

\textbf{Comparison.} Compared with Silo, \myarch uses fine-grained SSN to 
track indispensable dependencies among transactions, 
which achieves excellent scalability, while significantly shortening the commit 
latency of transactions. 
Compared with NVM-D and NV-Logging, \myarch uses full-time logger threads 
responsible for persisting log records. 
Due to effective batching mechanism (e.g. group commit~\cite{VLDB:hagmann1987reimplementing}), this can avoid flushing frequently, which 
significantly impacts performance 
when the database system is deployed on a server with slower IO devices. 
And this way also exhibits good performance for database systems deployed on 
NVM-based machines. 
Moreover, compared with \gsn of NVM-D, 
\mylsn does not track the WAR dependencies among transactions.
In other words, a transaction does not need to modify the special 
field (i.e., \mylsn) of each tuple read by the transaction itself. 
Consequently, \myarch 
exhibits higher performance in hybrid workloads.


%% file: logging.tex
\renewcommand{\mylsn}{SSN\xspace}
\renewcommand{\mylsns}{SSNs\xspace}
\newcommand{\durablessn}{DSN\xspace}
\newcommand{\durablessns}{DSNs\xspace}
\newcommand{\committedssn}{CSN\xspace}
\newcommand{\readset}{\ensuremath{\textit{RS}}\xspace}
\newcommand{\writeset}{\ensuremath{\textit{WS}}\xspace}
\newcommand{\segindex}{\ensuremath{\textit{S}}\xspace}
\newcommand{\segssn}{\textit{ssn}\xspace}
\newcommand{\segallocbyte}{\textit{allocated\_bytes}\xspace}
\newcommand{\segbufferedbyte}{\textit{buffered\_bytes}\xspace}
\newcommand{\segstat}{\textit{stat}\xspace}
\newcommand{\segstartoff}{\textit{start\_offset}\xspace}
\newcommand{\curwriteseg}{\textit{cur\_generate\_seg}\xspace}
\newcommand{\curflushseg}{\textit{cur\_flush\_seg}\xspace}
\newcommand{\hangbytes}{\textit{hang\_bytes}\xspace}
\newcommand{\offset}{\textit{offset}\xspace}
\newcommand{\lbs}{LBS\xspace}
\newcommand{\qww}{\textit{Qww}\xspace}
\newcommand{\qwr}{\textit{Qwr}\xspace}
\newcommand{\buffera}{\textit{LA}\xspace}
\newcommand{\bufferb}{\textit{LB}\xspace}
\newcommand{\trackingtable}{\textit{TT}\xspace}
\newcommand{\wno}{\textit{wno}\xspace}
\newcommand{\bssn}{\textit{bssn}\xspace}
\newcommand{\blsn}{\textit{blsn}\xspace}
\newcommand{\dlsn}{\textit{dlsn}\xspace}
\newcommand{\seg}{\textit{seg}\xspace}
\newcommand{\ssn}{\ensuremath{\textit{SSN}}\xspace}
\newcommand{\csn}{\ensuremath{\textit{CSN}}\xspace}

\section{Logging for Recoverability}\label{sec:paralogging}
In this section, we describe detailed designs of our logging approach named \myarch. We first provide an overview  
and then present how we address technical challenges to achieve scalable and parallel logging with guaranteed atomicity and durability of transactions. 

\begin{figure*}[!htpb]
	\centerline{\includegraphics[scale=.65]{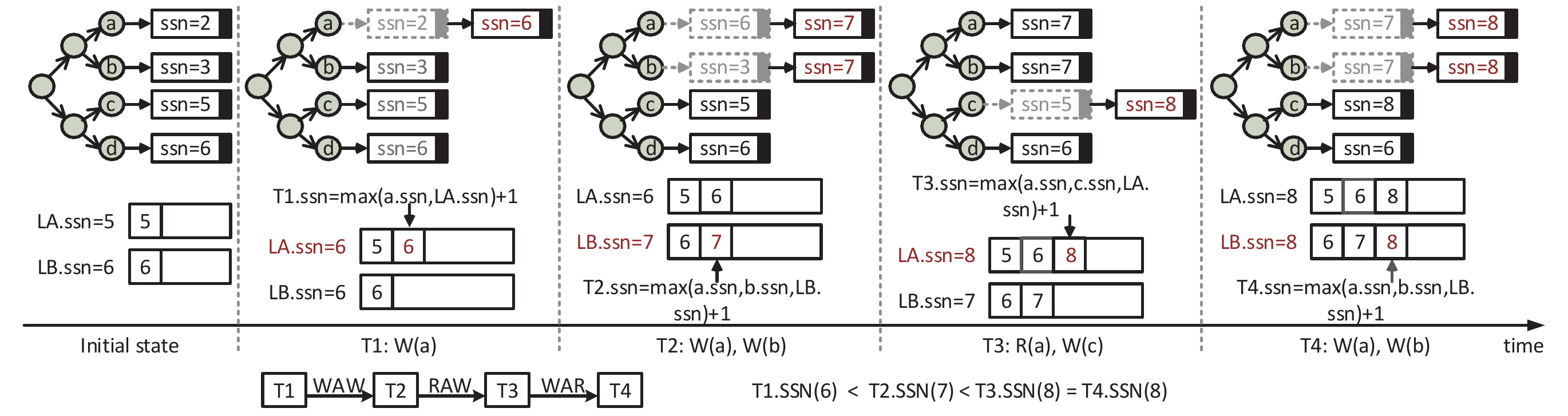}}\vspace{-2mm}
	\caption{An example of four transaction calculating \mylsns in a decentralized manner}
	\label{fig:calculatessn}\vspace{-3mm}
\end{figure*}

\subsection{Overview}\label{subsec:paral_arch}
The architecture of our logging \myarch is illustrated in Figure~\ref{fig:loggingoverview}. Multiple log buffers are utilized  to cache transaction updates in the form of log records. It includes two types of threads: worker threads and logger threads. There is a one-to-one mapping between logger threads and log buffers 
and a many-to-one mapping between worker threads and log buffers. Each worker thread produces its log records and copies them into its mapped log buffer.  In Figure~\ref{fig:loggingoverview}  there are two log buffers $\buffera$ and $\bufferb$, two logger threads and four worker threads. Worker thread $1$, $3$ are mapped to \buffera and worker thread $2$, $4$ are mapped to \bufferb. Each worker thread has two private commit queues (\qww and \qwr), 
the \qww is used to commit transactions only contain WAW dependencies and the \qwr is used for transactions with potential RAW dependencies.
The whole process of logging is based on a three-staged logging pipeline:  a prepare stage, a persistence stage and a commit stage.

In the prepare stage, each worker thread independently generates a log record for its own transaction and writes it into its mapped log buffer. Before copying the log record, each worker thread allocates a scalable sequence number (\mylsn) to the transaction and log record. The \mylsn provides a partial order, 
which tracks the RAW and WAW dependencies among transactions. 
When log record has been filled, the transaction is pushed to a commit queue and waits to be committed. If it is a read-only or read-write transaction, the worker thread push it into \qwr. If the transaction only contains write operations, it is pushed into \qww. 
As shown in Figure~\ref{fig:loggingoverview}, since all transactions' log records have been cached in their log buffers, 
$\txn_1$ with \mylsn$=6$, $\txn_2$ with \mylsn$=7$ and $\txn_4$ with \mylsn$=8$ are pushed into their own \qww, 
and $\txn_3$ with \mylsn$=8$ is pushed into its \qwr. 
In this stage, all steps are completely lock-free and can be done concurrently.



In the persistence stage, each logger thread independently flushes buffered log records into a bound storage device and advances its own durable \mylsn (denoted by \durablessn). 
That is to say, each logger thread or log buffer owns a \durablessn, and a \durablessn is the most recently flushed \mylsn, which is a durability indicator of log records for that log buffer. 
As illustrated in Figure~\ref{fig:loggingoverview}, log buffer $\buffera$'s \durablessn is $6$ and log buffer $\bufferb$'s \durablessn is $8$. 
Due to concurrent \mylsn allocation and memory copying in the prepare stage, 
it inevitably creates buffer slots which have been occupied by transactions but not yet completely filled in each log buffer.  If log records with buffer slots become persistent, the database system cannot be restored to a correct state. 
To prevent the fatal error, each logger thread tracks buffer slots and advances its \durablessn by only forcing sequentially buffered log records to storage devices with a segment-based approach.

In the commit stage, \myarch needs to determine which transactions can be committed. 
According to recoverability defined in Section~\ref{subsec:difinition}, 
a transaction can be committed if 
($1$) its log record has been persistent; ($2$)  the log records of 
its RAW predecessors have been persistent.
Transactions with RAW dependencies are committable only if they satisfy the 
condition ($1$) and condition ($2$) at the same time; other transactions can be committed  
when they meet the condition ($1$). 
Checking the first condition is straightforward by simply comparing the assigned \mylsn of the transaction with the \durablessn of the corresponding log buffer. However, checking the second condition is nontrivial. 
That is because log records from transactions with RAW dependencies may be distributed in different log buffers. 
Hence, we calculate a global committable \mylsn (denoted as \committedssn).  
The \committedssn is the committability indicator for transactions with RAW dependencies. When entering commit stage, each worker thread first checks its \qww and commits transactions whose \mylsns are smaller than \durablessn of their mapped log buffer, 
then checks the \qwr and commits transactions whose \mylsns $\leq$ \committedssn. 
As shown in Figure~\ref{fig:loggingoverview}, the \committedssn is equal to $6$. 
$\txn_3$ in \qwr can not be committed because its \mylsn$=8$ is larger than \committedssn. 
But $\txn_1$, $\txn_2$ and $\txn_4$ in \qww can be committed as their log records have been persistent.

\subsection{Scalable Sequence Number}\label{subsec:paral_psn}
We only need to focus on the WAW and RAW dependencies among transactions according to the definition of recoverability. 
For WAW dependency, if a transaction $T_j$ is WAW dependent on $T_i$, then the assigned log sequence number of $T_i$ should be less than that of $T_j$. To ensure this, we develop a scalable sequence number (\mylsn) in \myarch. 
Note that \mylsn tracks both the WAW and RAW dependencies among transactions. 
RAW dependency is necessary.  
Maintaining this dependency in \mylsn don't introduce additional overhead. 
Different from previous approaches, we do not calculate the sequence number using a centralized allocator~\cite{VLDB:johnson2010aether,VLDB:johnson2012scalability,VLDB:jung2017scalable,VLDB:tu2013speedy,VLDB:diaconu2013hekaton}. 
Instead, a transaction's \mylsn is allocated in a distributed manner based on its accessing tuples and the log buffer caching its log record. 
The distributed nature avoids the serialization bottleneck, making the algorithm highly scalable.

\textbf{Calculating \mylsn.} Our logging approach keeps \mylsns co-located with each log buffer and tuple.  
Each log buffer maintains the \mylsn of the log record that has been most recently cached in it and each tuple contains the \mylsn of the transaction that has most recently modified it. Algorithm~\ref{algo:acquiressn} shows the procedure how to calculate \mylsn for a transaction. 
Once a transaction $\txn$ is ready to write its log record, 
it first computes its \mylsn as the smallest number that is ($i$) larger than the \mylsn of any tuple read or written by the transaction (line 1-4) and 
($ii$) larger than the \mylsn of the log buffer serving the transaction (line 6-12). 
Next it advances the \mylsn of the log buffer to the transaction's \mylsn. Then the \mylsn is written into each tuple updated by the transaction (line 13-15). 
In addition to the \mylsn, 
we also calculate an offset for each log record,
which points the space of log buffer the log record will be cached. 
To address the contention on a log buffer, 
we use a compare-and-swap instruction to assign the \mylsn.  

\vspace{-2mm}
\input{acquiressn.tex}
\vspace{-2mm}


In particular, Figure~\ref{fig:calculatessn} illustrates the procedure of four transactions to calculate their \mylsns. 
As $\txn_1$ updates tuple $a$ and its log record will be cached in log buffer $\buffera$, 
it gets the \mylsn of tuple $a$ ($a.ssn=2$) and the \mylsn of log buffer $\buffera$ ($LA.ssn=5$), 
and calculates its \mylsn as $max(a.ssn, LA.ssn)+1=6$. 
Then the result is used to update the \mylsn of log buffer $\buffera$ and is stored in tuple $a$. 
For the $\txn_2$ that overwrites $\txn_1$'s update in terms of tuple $a$, 
its \mylsn must track the WAW dependency between $\txn_1$ and $\txn_2$. 
When calculating the \mylsn, $\txn_2$ acquires the largest \mylsn among tuple $a$, $b$ and log buffer $\bufferb$, 
and then increments the value by one. 
As a result, the \mylsn of $\txn_2$ is $7$ that is larger than that of $\txn_1$. 
In the same way, $\txn_3$ computes its \mylsn as $8$, 
which ensures the RAW dependency between $\txn_2$ and $\txn_3$. 
Since \mylsn does not track the WAR dependency between transactions, 
$\txn_3$'s \mylsn is not written into tuple $a$ that is only read by $\txn_3$. 
Therefore, \mylsn of $\txn_4$ is $8$, which equals to that of its WAR predecessor $\txn_3$. 

\subsection{Commit Protocol}\label{subsec:paral_commitpro}
\begin{figure}[!t]
	\centerline{\includegraphics[scale=.65]{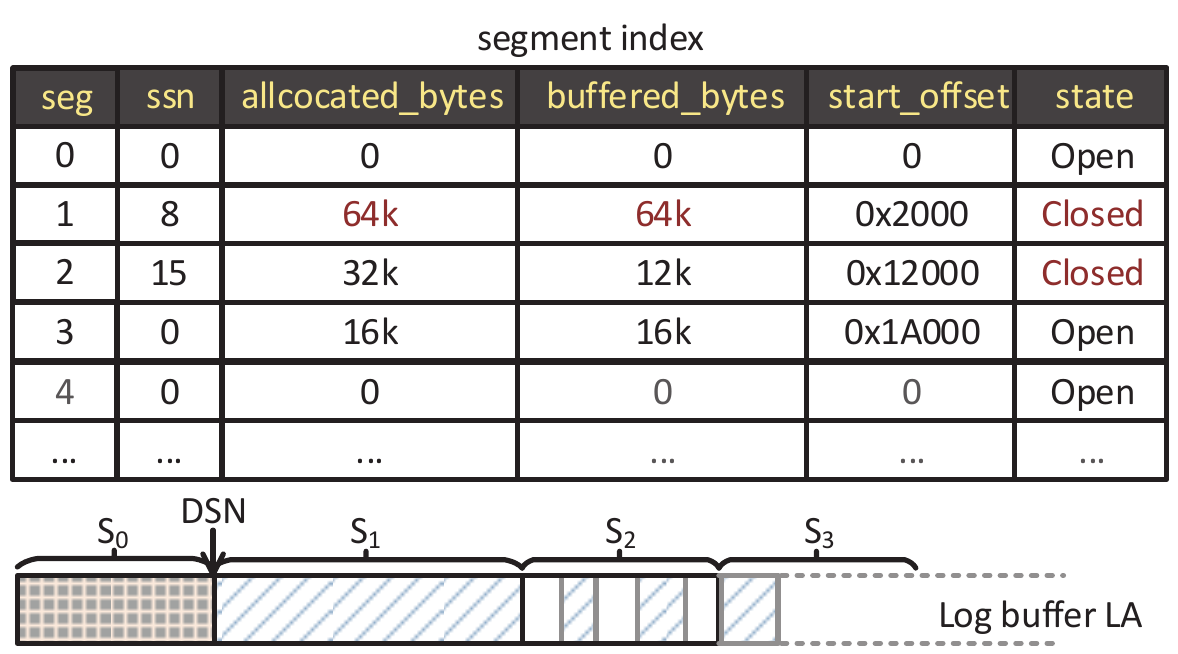}}\vspace{-2mm}
	\caption{Overview of a segment index.}
	\label{fig:advancedsn}\vspace{-2mm}
\end{figure}
\myarch proposes a fast commit protocol, 
which does not require sequential transaction committing.
In the commit state, 
transactions in queue \qww can be safely committed by comparing their \mylsns with corresponding \durablessns, 
while transactions in queue \qwr are committed by tracking \committedssn. \durablessns and \committedssn can be computed based on \mylsns.  
Next, we describe how to advance the \durablessn of each log buffer and compute the \committedssn in detail. 


\textbf{Advancing \durablessn.} 
To address the challenge caused by concurrent \mylsn allocating and log records copying, 
we use a segment-based approach to track buffer holes in each log buffer and advance the \durablessns. The segment-based approach logically divide each log buffer into segments by different size, and each segment is treated as a tracking or flushing unit. Only when all the log records in a segment have been sequentially buffered, 
the bound logger thread can force them into its stable storage device and advance the \durablessn to the largest \mylsn of these log records. 

 We maintain a table called a \emph{segment index} (or \emph{$\segindex$} in short) for each log buffer, as shown in Figure~\ref{fig:advancedsn}. Each entry of \segindex is a segment. Each segment is a quintuple of \emph{$<$\segssn,\segallocbyte, \segbufferedbyte, \segstartoff, \segstat$>$}, 
where the $\segssn$ is the largest \mylsn of log records in a segment, 
the \segallocbyte and \segbufferedbyte denote the cumulative byte count of allocated and buffered log records, 
the \segstartoff points the starting offset of a segment in its log buffer  
and the \segstat indicates the segment's status ($\segstat=closed$ means that a segment has been established).  
Each segment is generated in two different ways based on the gap between procedures (e.g. worker threads) and the consumer (e.g. logger thread): 
(1) a segment is established by a worker thread if it finds the cumulative byte count of allocated log records is larger than I/O unit size; 
(2) a segment is established by a logger thread when the time waiting for flushing exceeds the \emph{flush time} of the group commit strategy~\cite{VLDB:hagmann1987reimplementing}. 
We call the worker thread or logger thread generates segments a \emph{segment thread}. 
In Figure~\ref{fig:advancedsn}, $\segindex_3$ is the segment being generated, $\segindex_1$ and $\segindex_2$ are the segments have been established. 
In the generated segments, only $\segindex_1$ can be persisted but  $\segindex_2$ can not. 

As shown in Algorithm~\ref{algo:advancessn}, once the generating condition of segments is triggered, the segment thread establishes a segment by updating the \segssn and \segstat of the currently generating segment (pointed by a \curwriteseg), 
then atomically increments the \curwriteseg (line 2-7). 
Each logger thread periodically checks the state of segment that is about to be flushed (pointed by a \curflushseg). 
Only when the \segstat of the segment is closed and the \segallocbyte is equal to the \segbufferedbyte, 
the bound logger thread persists its log records and advances the \durablessn to the \segssn of the segment (line 10-13). 
Then it resets the segment to the initial state and atomically increments the \curflushseg (line 15-16). 

\input{advancessn.tex}

\textbf{Computing \committedssn.}  The \committedssn keeps track of the smallest value of \durablessns. 
Each logger thread reads the \durablessns of log buffers and advances the \committedssn to the smallest \durablessn, 
as described in line $19-21$ of Algorithm~\ref{algo:advancessn}. 
Transactions with RAW dependencies can be committed if their \mylsns are less than \committedssn. 
Consider a transaction $\txn_j$ is RAW dependent on $\txn_i$, 
$\txn_j$ can not be committed before $\txn_i$. 
Based on the calculation of their \mylsns, we always have $\ssn_i < \ssn_j$. 
Comparing their \mylsns and \committedssn, we can ensure they are committed in a correct order: 
(1) if $\ssn_i < \ssn_j \le \csn$, where both the log records of $T_i$ and $T_j$ have been persistent, then $T_i$ and $T_j$ are committable; 
(2) if $\ssn_i \le \csn < \ssn_j$, then only $\txn_i$ is allowed to be committed; 
(3) if $\csn < \ssn_i < \ssn_j$, where log record of $T_i$ is not persistent, 
so none of $\txn_i$ and $\txn_j$ can be committed. 


\subsection{Integrating with OCC} 
\myarch can be smoothly embedded into in-memory database systems with any concurrency control mechanism.  
It is particularly suitable for OCC.  
To ensure correctness in the presence of failure, 
a schedule should be serializable and strict~\cite{VLDB:bernstein1987concurrency}. 
Strictness reinforces a serializable schedule with additional constraints on the commit order of all transactions, 
which requires that the transaction committing tracks RAW and WAW dependencies among transactions. 
Therefore, we can directly use \mylsn instead of traditional algorithm's commit timestamp to guarantee the serializability and strictness. 
In \myarch, we implement a variant of optimistic concurrency control. 
We describe how to run transactions with \mylsn as follows. 
Under OCC, \myarch executes a transaction in three phases: 

During the read phase, transactions can access database without acquiring any lock. 
And each transaction maintains a private read set  and write set. 
Accessed tuples with their \mylsn are copied to the read set 
and modifies tuples with their new state are written into the write set. 

During the validation phase, \myarch checks whether a transaction can be committed based on the \mylsns stored in the transaction's read and write set, 
then computes a new \mylsn as the commit timestamp of the transaction. 
At first, a worker thread, on behalf of a transaction, locks all the tuples in write set in their primary key order. 
Using the fixed locking order avoids deadlocks 
with other committing transactions at the same time. 
This tips is also used in other OCC algorithms~\cite{VLDB:tu2013speedy,VLDB:yu2016tictoc}.
After all write locks are acquired, the worker thread begins to examine all the tuples in the read set.  
If (1) tuples in the read set are not locked by other transactions, 
and (2) the \mylsn of each tuple is not changed, 
the transaction is allowed to be committed. 
Otherwise, the worker thread releases all locks on the modified tuples and aborts the transaction. 
Once validation is successful, 
the transaction calculates a new \mylsn based on the allocation approach introduced earlier and enters the logging phase. 

During the write phase, 
modified tuples with the new \mylsn computed in the validation phase are written to the database. 
After commit, all write locks are released, making the changes visible to other transactions. 
Combined with early lock release, a transaction enters the write phase once it acquires the new \mylsn. 
Although incoming transactions can read the uncommitted transaction's dirty data, 
\myarch guarantees a consistent state as the transaction is actually committed before its RAW successors.

%% file: acquiressn.tex
\begin{algorithm}[]
\caption{Calculating SSN}\label{algo:acquiressn}
\KwIn{a transaction T,read set RS, write set WS}
\KwIn{a log buffer L, log size len}
\SetKwFunction{cas}{CAS}
\SetKwFunction{max}{max}
\SetKwFunction{barrier}{COMPILER\_BARRIER}
\SetKwFunction{faa}{FETCH\_ADD}
$base = 0$\\
\For{e in RS $\cup$ WS}{
	$base = \max{e.ssn, base}$\\
}
\eIf{WS is not empty}
{
\While{$\cas{L.latch, false, true}$}{
$T.ssn = \max{base, L.ssn} + 1$\\
$L.ssn = T.ssn$\\
$\faa{L.\offset, len}$\\
\barrier\\
$L.latch = false$\\
}
\For{e in WS}{
	$e.ssn = T.ssn$\\
}
}
{
$T.ssn = base$\\
}
\end{algorithm}

%% file: advancessn.tex
\begin{algorithm}[!htpb]
\caption{Advancing DSN and CSN}\label{algo:advancessn}
\KwData{segment index \segindex, logger thread Lg}
\KwData{log buffer L, log buffer set LBS}
\KwData{I/O unit size IO, flush timer FT}
\SetKwFunction{mod}{modulo}
\SetKwFunction{faa}{FETCH\_ADD}
\SetKwFunction{flush}{flush}
\SetKwFunction{establish}{\textrm{Establishing Segment}}
\SetKwFunction{advdsn}{\textrm{Advancing DSN}}
\SetKwFunction{advcsn}{\textrm{Advancing CSN}}

\SetKwFunction{min}{min}
\SetKwProg{Fn}{Procedure}{}{}
\Fn{\establish{}}{
$i = \curwriteseg$ \mod $\segindex.size$\\
\If{$(\segindex[i].\segallocbyte \geq IO \parallel$
$Lg.wait\_time \geq FT$) $\&\&$ $\segindex[i].\segstat$ $!= closed$}{
$\segindex[i].\segssn = L.ssn$\\
$\segindex[i].\segstat = closed$\\
$\segindex[i+1].\segstartoff = L.\offset$\\
$\faa{\curwriteseg,1}$\\
}
}
\Fn{\advdsn{}}{
$i = \curflushseg$ \mod $\segindex.size$\\
\While{($\segindex[i].\segallocbyte == \segindex[i].\segbufferedbyte$) $\&\&$ ($\segindex[i].\segstat == closed$)}{
\flush{$\segindex[i].\segstartoff, \segindex[i].\segallocbyte$}\\
$L.dsn = \segindex[i].\segssn$\\
\barrier\\
$\faa{\curflushseg,1}$\\
$\segindex[i].reset()$\\
}
}
\Fn{\advcsn{}}{
\For{l in LBS}{
	$csn = \min{csn, l.dsn}$\\
}
}

\end{algorithm}

%% file: recovery.tex
\newcommand{\recoveryssn}{RSN\xspace}
\newcommand{\rssnstart}{RSNs\xspace}
\newcommand{\rssnend}{RSNe\xspace}
\section{Recovery} 
\label{sec:recovery}
In this section, we present the recovery algorithm based on our parallel logging. We rely on both checkpoints and log files to restore the database system into a consistent state.

\textbf{Checkpoints.}  To accelerate recovery from a crash, in-memory database systems mandate periodic checkpoints of their state during logging process. 
The main challenge in checkpoint production is to produce checkpoints as quickly as possible without deteriorating logging throughput. 
As with parallel logging approach, \myarch uses multiple checkpoint threads to produce checkpoints and write concurrently them into multiple storage devices. 
There is a one-to-one mapping between checkpoint threads and storage devices. 
In our implementation, all tuples are evenly divided into multiple partitions and each partition is processed by a checkpoint thread. 
When starting checkpointing, a checkpoint deamon starts up $n$ checkpoint threads. 
Each checkpoint thread walks over its assigned partition in key order and writes them into $m$ checkpoint files, wherer $m \times n$ is the total number of checkpoint files for recovery, which can be configured as the the number of CPU cores. 
At the same time,the checkpoint deamon records current \committedssn as \recoveryssn, where \recoveryssn indicates the start time of the checkpointing, 
and writes it into a metadata files after the checkpointing has been completed. 

In \myarch, we allow concurrent transactions to continue their execution during a checkpoint period. 
Transactions do not coordinate with checkpoint threads except per-tuple locks 
so that checkpoint threads might observe an inconsistent snapshot. 
This is usually called fuzzing checkpointing~\cite{VLDB:li1993post}. 
To recover a consistent database state, it is necessary both to restore the checkpoints and to replay log records whose \mylsn is larger than the \recoveryssn stored in the checkpoint metadata. 
Furthermore, if an in-memory database system adopts the early lock release method,  each transaction can release locks before committing. 
Hence checkpoint threads can read the uncommitted transactions' dirty data, 
which results in incorrect checkpoints. 
To avoid this problem, each checkpoint thread computes the largest \mylsn of tuples it observed during performing checkpointing. 
And the checkpointing is considered successful only if the latest \committedssn is larger than the calculated results of all checkpoint threads. 

\textbf{Failure recovery.} 
There are two main stages to restore the database state after a crash: 

In checkpoints recovery stage, a recovery deamon gets the \recoveryssn from the newest checkpoint metadata (named as \rssnstart), where the \rssnstart denotes the starting point for log recovery. 
Then the recovery deamon starts up $m\times n$ recovery threads to recover all checkpoints in parallel. 

In log recovery stage, recovery threads can concurrently replay the persistent log records by applying the last-writer-wins rule~\cite{VLDB:thomas1979lww}. 
As a durable log record may come from an uncommitted transaction, 
it is necessary to determine which persistent log records can be recovered. 
At the beginning of log recovery stage, the recovery deamon checks the most recently durable log record on each storage device 
and calculates the smallest \mylsn as \rssnend, 
where \rssnend is the ending point for log recovery. 
Transactions whose \mylsn $\leq$ \rssnend can be guaranteed to be committed. 
As a result, log records contains \mylsns are in $($\rssnstart, \rssnend$]$ can be safely recovered. 
In addition, transactions only contain WAW dependencies are committed when their log records have been durable. 
For log records of those transactions, 
they can be replayed as long as they are persistent, 
regardless of whether their \mylsns are larger than the \rssnend. 
To reduce the overhead of computing \rssnend, we shorten moderately the size of each log files. 
The recovery deamon calculates the \rssnend based on the latest log files, 
while recovery threads can concurrently replay log records from other log files.

%% file: evaluation.tex
\section{Evaluation}\label{sec:expr}
In this section, we present the experimental results of \myarch, 
which is implemented on the open-source codebase DBx1000~\cite{VLDB:DBx1000}. 
DBx1000 is an in-memory DBMS prototype that stores all database in main memory and provides different concurrency control protocols. 
We implement \myarch with the optimistic concurrency control of Silo and compare it with traditional approaches, 
and confirm the following: 
\begin{itemize}
\item With the decentralized \mylsn management policy, \myarch maintains the highest throughput compared with other transaction logging approaches.
 
\item The commit protocol requires only transactions with RAW dependencies to be committed in serial order gives super commit latency of transactions. 

\item \myarch breaks thoroughly the sequential constraints of traditional logging, making logging and recovery highly scalable. 
\end{itemize}

\subsection{Experimental Setup}
\label{subsec:setup}
\textbf{Hardware.} 
All experiments are executed on a server with 256GB DRAM 
and two Intel Xeon E5-2630 v4 processors clocked at 2.20GHz (20 physical cores in total). To reduce irrelevant lock contention, we run all experiments with hyperthreading being disabled. 
The server contains four pieces of PCIe SSDs of $1.6$ TB, each of which provides the peak sequential write throughput of $1.2$ GB/s with $21.5 us$ delay for the sequential writes of $16$ KB block. 

\textbf{NVRAM emulation.} The non-volatile memory module available on the market is designed to be integrated into NVDIMM enabled servers via DDR4 DIMM sockets~\cite{VLDB:agigarnvdimm,VLDB:vikingnvdimm}. The NVDIMM provides a nearly latency of DRAM at runtime. So we use DRAM in our experiments to emulate the performance of NVM.  We first estimate the data persistence costs of NVM based on Intel Labs' 
persistent memory evaluation platform (PMEP)~\cite{VLDB:zhang2015study} that 
configured with \texttt{CFLUSH} instruction,  
then configure the NVM latency to be 2$\times$ that of DRAM. 
The extra latency is added through a loop that busy-waits for a specified number of CPU cycles. 

\textbf{Variants.} 
We compare the performance of \myarch with other approaches in end-to-end experiments: 

1) \centralized. The traditional logging uses a central log buffer and a single IO device to store transactions' log records. 
Each transaction acquires its \lsn using a \texttt{fetch-add} atomic instruction and is committed in the total \lsn order.

2) \nvmd. The distributed logging~\cite{VLDB:wang2014scalable} is based on emulated non-volatile memory (NVM), where worker threads directly force their log records to NVM. Each transaction calculates its \gsn in a decentralized manner and is committed rigorously in the  \gsn order. 
The \gsn tracks all dependencies among transactions. 

3) \silo. The logging approach uses multiple log buffers bound with multiple IO 
devices to cache log records. 
All transactions are committed sequentially in the periodically-updated epoch order (epoch 
increments every $50$ ms).

4) \para. Our logging uses \mylsn to track WAW and RAW dependencies among transactions, and only transactions with RAW dependencies are committed in serial order. 


All variants are implemented in the open sourced database DBx1000 with the optimistic concurrency control of Silo for a fair comparison. By default, we run each variant with 2 SSDs, except for \centralized. The number of SSDs is equal to the number of log buffers and logger threads. We use group commit for each logger thread which maintains a log buffer ($30$ MB) to cache log records. Log records are flushed every $5$ ms or if the buffer is half full, whichever happens first. For \nvmd, each worker thread maps to exactly one SSD and flushes log records into its own log files.  For experiments with emulated NVM, we set the log buffer size to $1$ MB and flush the log every $5$ ms or the buffer is $1/10$ full.

\subsection{Workloads}
\label{subsec:exprworkload}
We use two benchmarks YCSB and TPC-C in our experiments to evaluate all variants. 

\textbf{YCSB.} 
The Yahoo! Cloud Serving Benchmark (YCSB) is representative of large-scale on-line services~\cite{VLDB:cooper2010benchmarking}. 
It has a single table with a primary key and 10 additional columns with 100 bytes. 
The dataset in the table is initialized to 10 million records. 
In our experiments, we mainly adopt two different kinds of workloads: 

1) \textbf{Write-only workload}. In this setting, there are only write 
transactions, each one updates all columns of one tuple accessed by 
the key value. 

2) \textbf{Hybrid workload}. 
In this workload, each transaction consists of a write operation updating one column of a tuple, 
and a key-range scan operation with a fixed scan length. 
By default, the key accessed by each transaction follows uniformly random 
distribution. 
This way can eliminate the effect of other contention (e.g., concurrency 
control) and focus on the evaluation of the logging approach. 
For each transaction, we run $10$M transactions five times and the results are averaged over the five runnings. 
 
\textbf{TPC-C.} 
This is a current industry-standard for OLTP applications that simulates a warehouse-centric order processing system~\cite{TPCC}. 
It contains nine tables and five transactions that consist of three read-write transactions and two read-only transactions. 
We only model two  transactions ( 50 \% Payment and 50 \% NewOrder) in our experiments. 
For each experiment, we populate the database with 20 warehouse and run $10$M transactions five times.

\subsection{Logging Performance}
\label{subsec:exprlogging}
In this section, we focus on evaluating the logging performance of all the variants. 
We first measure the throughput, commit latency, I/O bandwidth, runtime breakdown and scalability 
under YCSB with write-only workload and TPC-C. 
Then we further analyze the impact of commit protocol of all logging approaches under YCSB with hybrid workload.

\begin{figure}[!tpb]
	\centering
	\subfigure[
	YCSB \label{fig:log-tps-ycsb}]{\includegraphics[scale=.52]{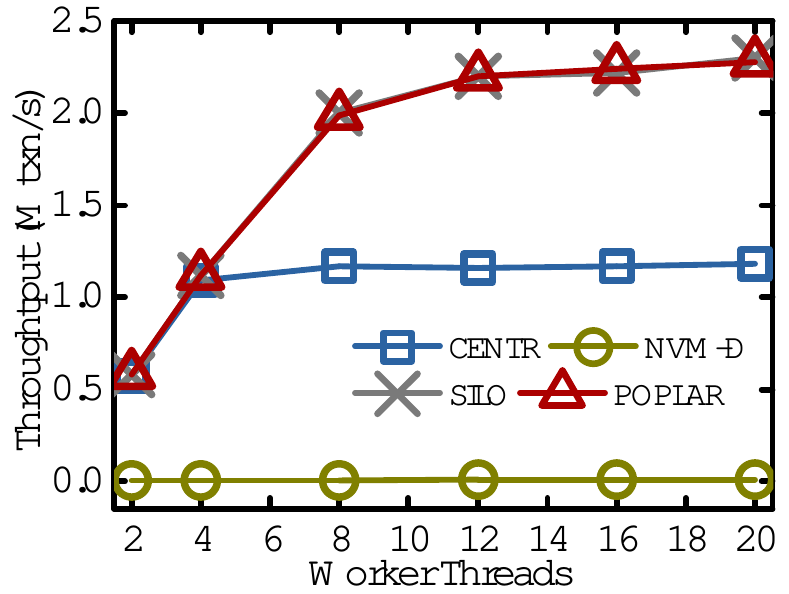}}
	\subfigure[TPC-C\label{fig:log-tps-tpcc}]{\includegraphics[scale=.52]{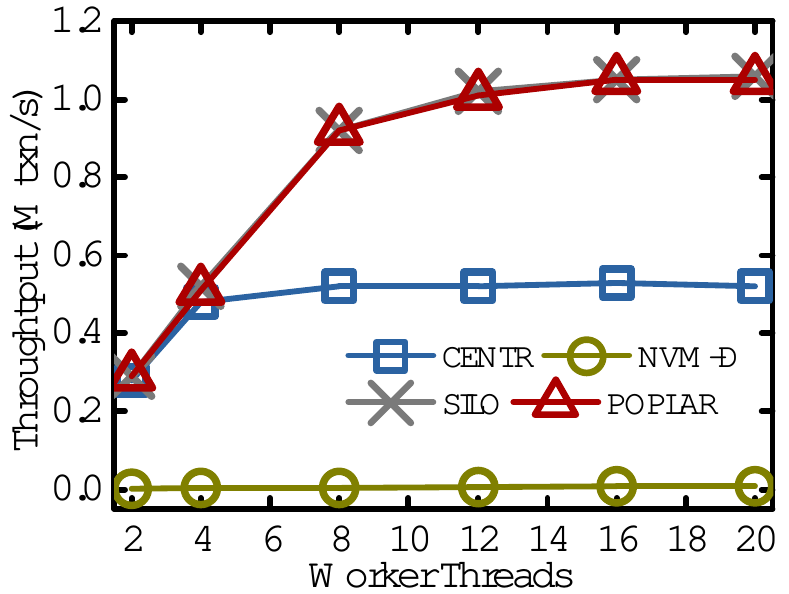}}
	\caption{Throughput. \para and \silo keep the highest performance by eliminating the IO bottleneck of \centralized.} 
	\label{fig::log-tps}
\end{figure}

\textbf{Throughput.} 
Figure~\ref{fig::log-tps} shows the throughput results. 
In this experiment, we use two pieces of SSDs to store log records
and vary the number of worker threads to observe the peak throughput of each approach. 
For the YCSB workload, as the number of worker threads increases, the 
throughput of each method rises 
steadily at first, but eventually reaches a saturation value. 
This is because the IO bandwidth of SSDs becomes the primary bottleneck when the volume of transactions grows. 
Using a single SSD as the permanent storage, \centralized shows the lowest throughput due to its limited IO bandwidth. 
Owing to the decentralized \mylsn allocation and multiple SSDs, 
\para exhibits the same excellent throughput as \silo, 
which improves near $2\times$ better performance than \centralized. 
For \nvmd, although the value of throughput increases with the number of worker 
threads, 
it can not rival the performance of other approaches at the same number of worker threads, 
where there is a wide performance gap ($\sim 280\times$) between \para and \nvmd. 
The worse throughput is caused by the fact that worker threads of \nvmd write log records directly into stable storages, which produces additional scheduling and is not suitable for SSDs and HDDs. 
For the TPC-C workload, 
the throughput of all variants grows almost linearly at first due to the scalable design for \lsn, \gsn and \mylsn, 
and eventually reaches a steady state because of the limited I/O bandwidth. 
Similar to the YCSB workload, 
\para and \silo show the highest throughput, which delivers $2\times$ better 
performance than \centralized and yields a $\sim131\times$ gain over \nvmd. 

\begin{figure}[!tpb]
	\centering
	\subfigure[
	YCSB\label{fig:log-io-ycsb}]{\includegraphics[scale=.52]{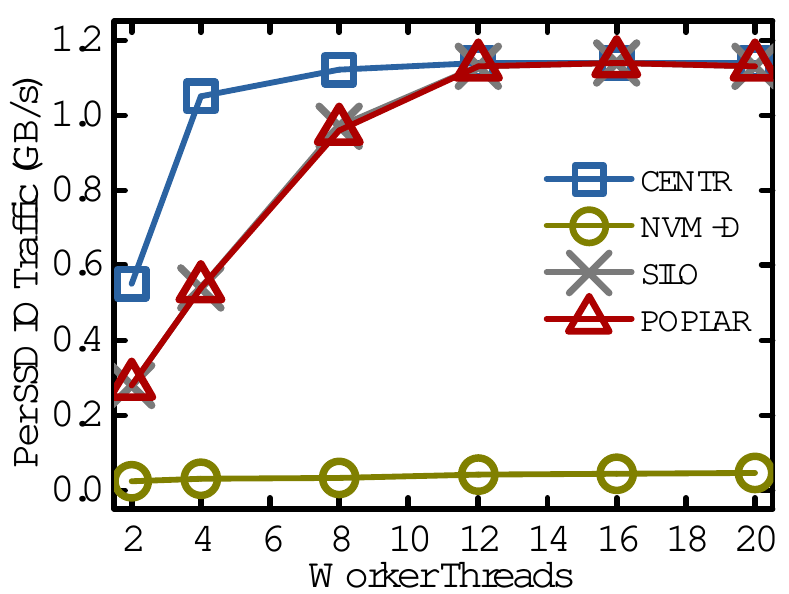}}
	\subfigure[TPC-C\label{fig:log-io-tpcc}]{\includegraphics[scale=.52]{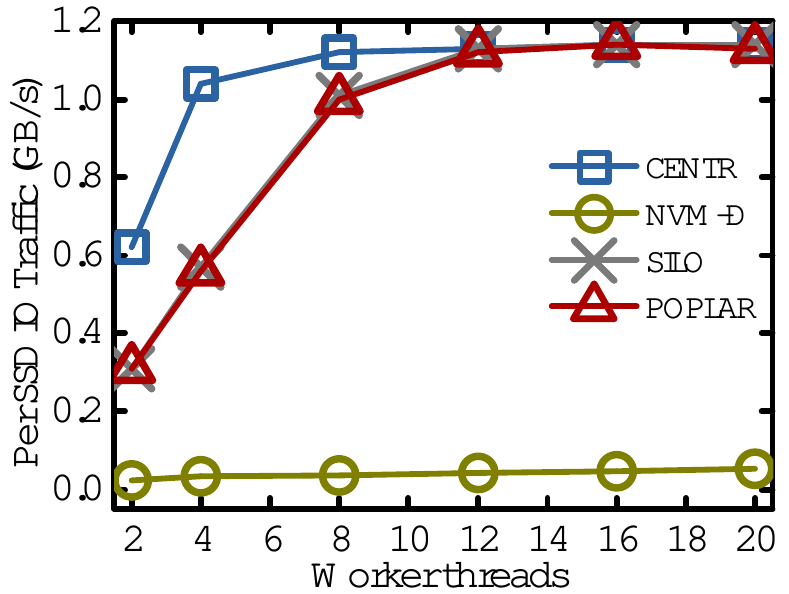}}
	\caption{Per Device IO Bandwidth. The limited IO bandwidth becomes the primary bottleneck of logging. Performance can further increase given more SSDs, except for \centralized.} 
	\label{fig::log-disk} \vspace{-3mm}
\end{figure}

\textbf{I/O bandwidth.} 
Next, we employ \textsf{nmon} to monitor the I/O bandwidth of each SSD used in all logging approaches. 
Figure~\ref{fig::log-disk} shows that for both YCSB and TPC-C, as the number of worker threads increases, 
the used I/O bandwidth of each method gradually grows and finally saturates 
the peak value of each SSDs. 
This demonstrates that limited I/O bandwidth is the primary bottleneck of logging. 
Therefore, \para and \silo have better throughput than \centralized by leveraging two SSDs as shown in Figure~\ref{fig::log-tps}
and their peak throughput can further increase given more I/O devices. 

\begin{figure}[!tpb]
	\centering
	\subfigure[
	YCSB\label{fig:log-latency-ycsb}]{\includegraphics[scale=.52]{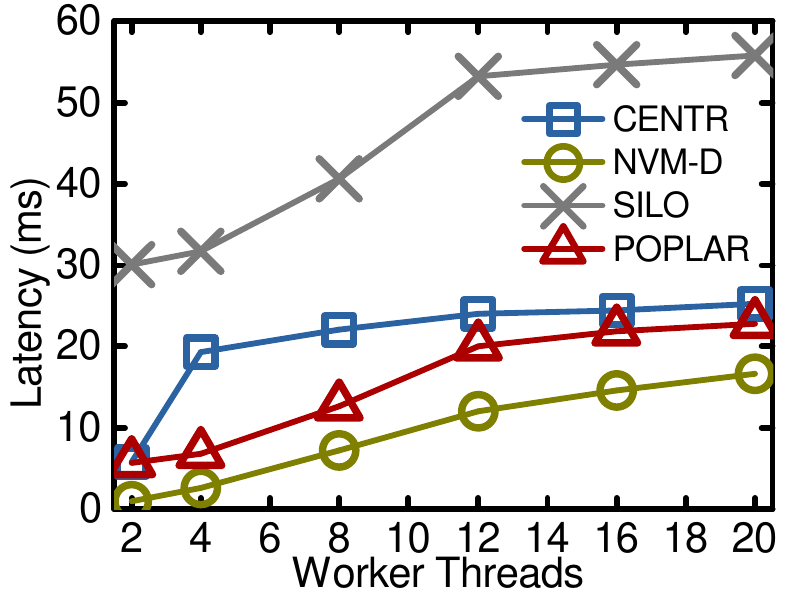}}
	\subfigure[TPC-C\label{fig:log-latebct-tpcc}]{\includegraphics[scale=.52]{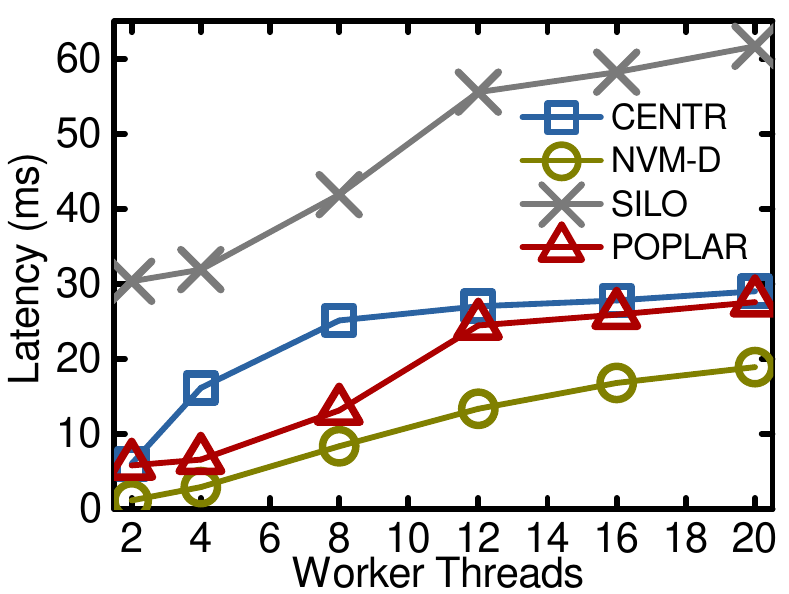}}
	\caption{Commit Latency. \silo has the longest commit latency all the time due to its epoch-based commit protocol.} 
	\label{fig::log-latency}  
\end{figure}

\textbf{Commit latency.} 
Figure~\ref{fig::log-latency} shows the commit latency of transactions in all logging approaches. 
We measure the commit latency by varying the number of worker threads. 
In all workloads, \silo has the longest latency ($\sim$ $6\times$) than that of other approaches all the time. 
Such poor performance of \silo is primarily 
owing to the epoch-based commit protocol demands 
a transaction can not be committed until log records of all transaction within the same epoch have been persisted. 
For \para and \centralized, they have short latency at low worker thread count, 
which stays around $5$ ms that is the group commit time interval. 
As the number of worker threads increases, the commit latency continues to increase 
due to the growth of flushing time. 
When the thread count is $20$, where \para and \centralized both saturate the I/O bandwidth, 
they have longer latency because a large number of transactions are waiting for flush and worker threads are blocked due the insufficient buffer space. 
Overall, \para outperforms \centralized and the commit latency improvement that \para contributes at low worker thread count is more than $2.5\times$. 
For \nvmd, the commit latency gradually increases in the graph. 
This is because more and more worker threads write log records to the same SSD. 

\begin{figure}[!tpb]
	\centering
	\subfigure[
	YCSB\label{fig:log-breakdown-ycsb}]{\includegraphics[scale=.52]{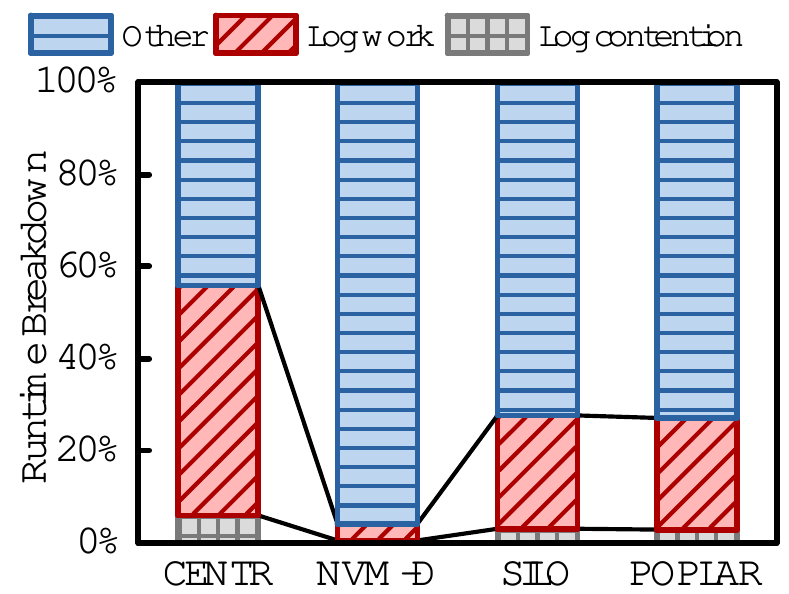}}
	\subfigure[TPC-C\label{fig:log-breakdown-tpcc}]{\includegraphics[scale=.52]{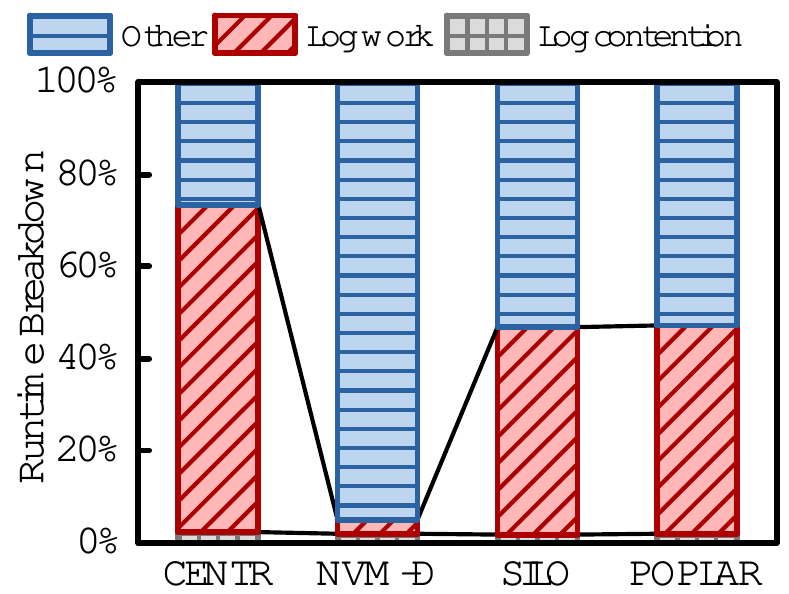}}
	\caption{Runtime Breakdown($20$ worker threads). Log contention consumes the least CPU time due to decentralized \mylsn allocation.} 
	\label{fig::log-breakdown}\vspace{-3mm} 
\end{figure}

\textbf{Runtime breakdown.} 
To investigate the effectiveness of \mylsn allocation, 
we collect the execution time of $20$ worker threads 
and turn them into time breakdown to identify the CPU cycles 
consumed by Log contention, Log work and Other. 
Log contention indicates sequence number (\lsn, \gsn and \mylsn) allocation, 
Log work includes log records inserting and waiting log buffer available, 
Other expresses the actual transaction logic and concurrency control. 
As shown in Figure~\ref{fig::log-breakdown}, both in YCSB and TPC-C, 
Log contention of all variants consumes the least CPU time due to their highly scalable sequence number algorithms. 
The higher CPU time of Log work in \centralized, \silo and \para is caused by the I/O bandwidth saturation. 
When the number of worker threads is $20$, each log buffer is quickly filled up 
which makes all worker threads to wait for available buffer space. 
This significantly increases the CPU execution cycles. 
It is obvious that the overhead of Log work can be reduced by adding I/O devices. 
Therefore, the CPU cycles of Log work of \para and \silo are less than \centralized's. 

\begin{figure}[!tpb]
	\centering
	\subfigure[
	YCSB\label{fig:log-scal-ycsb}]{\includegraphics[scale=.52]{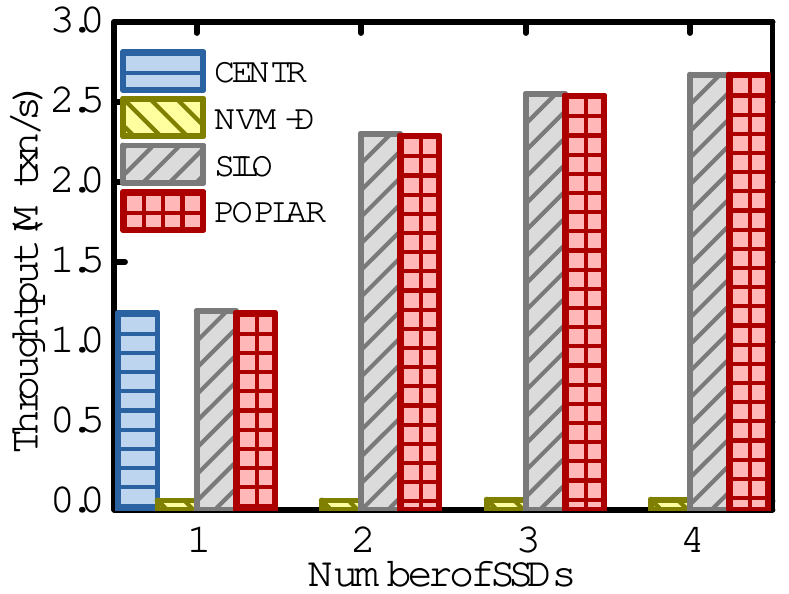}}
	\subfigure[TPC-C\label{fig:log-scal-tpcc}]{\includegraphics[scale=.52]{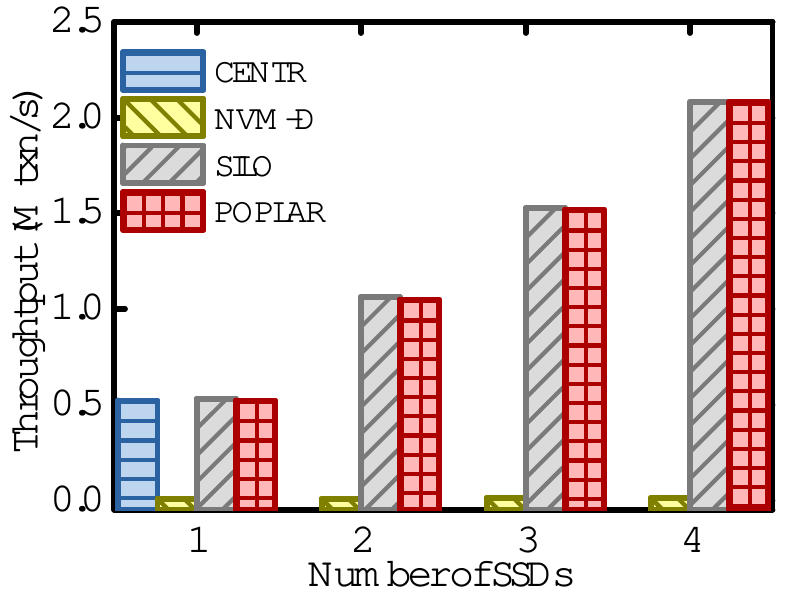}}
	\caption{Scalability of logging. The throughput of \para and \silo scales well as the number of SSDs increase while \centralized cannot.} 
	\label{fig::log-scal}  
\end{figure}

\textbf{Scalability.} 
Next, we measure the peak throughput to compare the scalability of different logging approaches by varying the number of SSDs. 
The experimental results of YCSB and TPC-C workloads are presented in Figure~\ref{fig::log-scal}. 
When using a single SSD, the throughput of \para and \silo is almost the same as that of \centralized. 
As the number of SSDs increases, more transactions can be flushed in parallel and more I/O bandwidth can be used, 
which makes the \para and \silo scale well, except for \centralized.
In the YCSB workload, when the number of SSDs is larger than $2$, the throughput of \para and \silo does not  
keep increasing as that in the TPC-C workload. 
Obviously, this is caused by the limited CPU processing power. 
If we give more cores in each processor, the performance can scale well as in 
the TPC-C. 
The performance of \nvmd also grows when we increase the number of SSDs. 
But its peak throughput is seriously lower than \para and 
\silo in all experiments due to its design based on NVM, 
which is not suitable for common I/O devices (e.g., SSDs).

\begin{figure}[!tpb]
	\centering
	\subfigure[
	Throughput\label{fig:nvm-tps}]{\includegraphics[scale=.52]{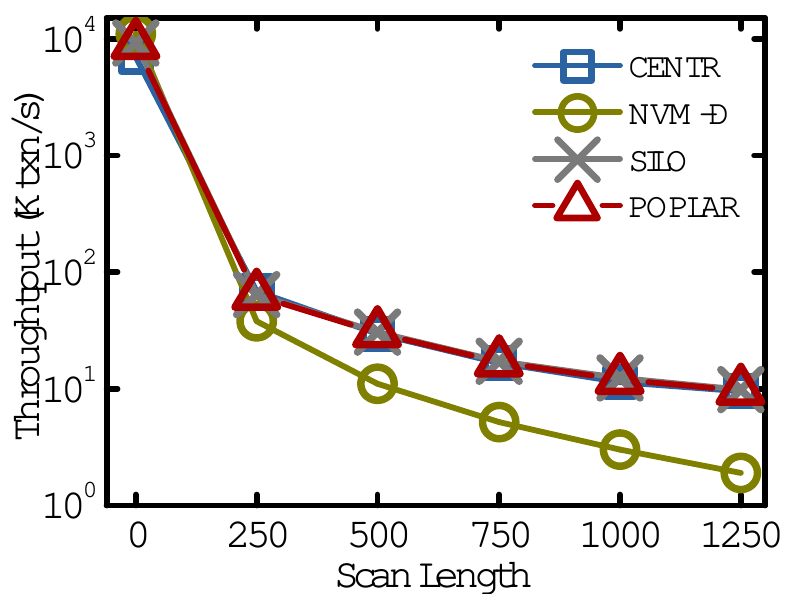}}
	\subfigure[Latency\label{fig:nvm-latency}]{\includegraphics[scale=.52]{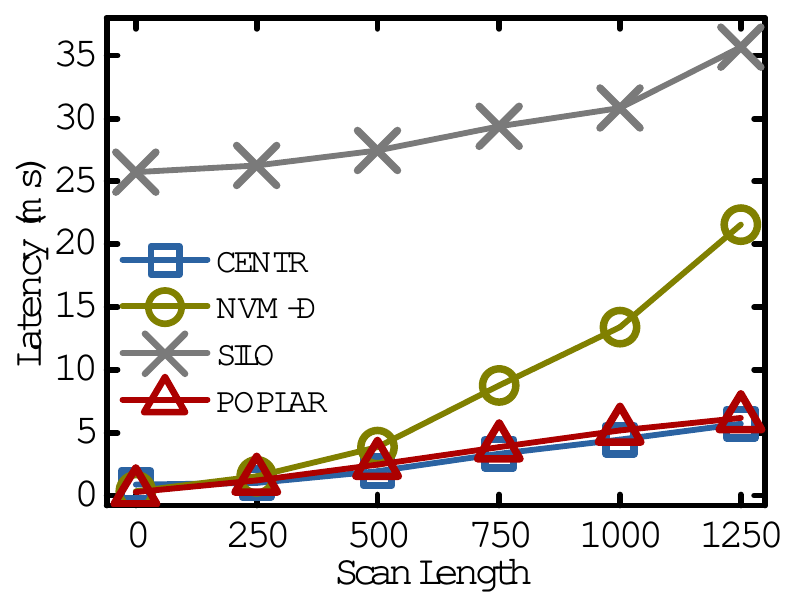}}
	\caption{Impact of the commit protocol. \para maintains the highest throughput and super commit latency on NVM devices.} 
	\label{fig::nvm} \vspace{-3mm} 
\end{figure}

\textbf{Commit protocol with NVM emulation.} 
To further investigate the impact of different commit protocols in logging approaches, 
we use YCSB with the hybrid workload in this experiment. 
The hybrid workload consists of a write query and a scan query with a fixed scan length. 
The scan length is used to indicate how many tuples might be accessed, 
which implies the proportion of RAW and WAR dependencies. 
In order to evaluate the impact of read operation on different commit protocols, 
we measure the performance of all variants by varying the scan length. 
All experiments are run on the emulated NVM that eliminates the effect of IO bandwidth on experiment results. 
In this experiment, we fix the number of worker threads to $20$ and the number of logger threads of \para and \silo to $2$. 

Figure~\ref{fig::nvm} shows the performance in terms of throughput and commit latency of all variants. 
When the scan length equals to $0$, all logging approaches have almost the same throughput, but \silo gets the commit latency of $25.7$ ms, whereas others have about $0.23$ ms.  
This huge performance gap ($\sim 112\times$) between \silo and other methods is mainly due to the epoch-based commit protocol used in \silo where transactions spend a lot of time waiting to be committed.  As the scan length increases, 
the performance of all logging approaches declines. 
This is because worker threads need to handle increasing complex transaction logic when the scan length is larger. 
And when the scan length is larger than $0$, \nvmd exhibits the worse throughput and logger commit latency than that of \para. 
\nvmd requires that transactions with WAW, WAR or RAW dependencies must be committed in order, which increases the overhead of  sequence number (\gsn) allocation and extends unnecessary waiting time. 
The cost of \gsn allocation is linearly proportional to the number of accessed tuples. 
Therefore, the throughput of \nvmd reduces linearly and commit latency grows linearly as the scan length increases. 
However, \para only needs to track the WAW and RAW dependencies among transactions and its \mylsn allocation is not affected by the scan length. 
Hence, \para maintains the super throughput and excellent commit latency. In this picture, \centralized also has a high performance due to the absent of any bottlenecks in our experiment environment. 
Obviously, if the processor has more cores (hundreds of thousands), 
contention over the centralized log buffer will become the primary bottleneck, whereas the contention can be addressed by increasing the number of log buffers in \para. 

\subsection{Recovery Performance}

\begin{table}[!tpb]
	\centering
	\caption{Recovery performance for YCSB.} \label{tab:2}
	\begin{tabular}{|p{3.8cm}|p{1.1cm}|p{0.8cm}|p{1.1cm}|}\hline
		&\centralized &\silo &\para \\ \hline
		Checkpoint Recovery Time&10.91s&5.34s&5.35s \\ \hline
		Log Recovery Time&93.25s&45.73s&45.75s \\ \hline
		Total Time&104.16s&51.07s&51.10s\\ \hline
	\end{tabular}
\end{table}
\begin{table}[!tpb]
	\centering
	\caption{Recovery performance for TPCC.} \label{tab:3}
	\begin{tabular}{|p{3.8cm}|p{1.1cm}|p{0.8cm}|p{1.1cm}|}\hline
		&\centralized &\silo &\para \\ \hline
		Checkpoint Recovery Time&50.19s&24.25s&24.29s \\ \hline
		Log Recovery Time&146.81s&70.93s&71.34s \\ \hline
		Total Time&197.00s&95.18s&95.63s\\ \hline
	\end{tabular}
\end{table}

We now evaluate the recovery performance of \centralized, \silo and \para. 
We first measure the recovery time and then explore the scalability of the recovery subsystem in different logging approaches. 
We run YCSB with write-only workload and TPC-C 
and in each case, the number of recovery threads is fixed to $20$.

\textbf{Recovery time.} 
Table~\ref{tab:2} and Table~\ref{tab:3} show the recovery time of YCSB and 
TPC-C in detail. 
In this experiment, we use two pieces of SSDs to store checkpoint files and log files, except for \centralized. 
For YCSB workload, all logging approaches must recover $9$ GB of checkpoints and $77$ GB of logs to recreate a database. 
As shown in Table~\ref{tab:2}, the \centralized has the largest recovery time compared with the \para and \silo.
This is because in \centralized,  only one recovery thread can load checkpoints and log files from the single I/O device at the same time, although recovery threads are able to  concurrently replay the checkpoints and logs in memory. 
Therefore, the limited single I/O bandwidth seriously reduced the recovery performance. 
Owing to the parallel load, the \para and \silo take less time ($\sim 51.10$ s) to recovery, 
which achieves more than $2.1\times$ better performance than \centralized. 
For TPC-C workload, there are a $40$ GB of checkpoints and $117$ GB of log files used to recovery. 
Table~\ref{tab:3} shows the results that represent the same trends as the YCSB workload. 
And \para takes about $95.63$ s to recovery the checkpoints and log files. 
In this experiment, it can be seen that recovery time is proportional to the amount of data that must be read to recover, 
and log recovery is the limiting factor in recovery. 
Thus, a reasonable decision to checkpoint frequently can effectively shorten recovery time without affecting normal transaction execution.

\begin{figure}[!tpb]
	\centering
	\subfigure[YCSB\label{fig:recovery-scal-ycsb}]{\includegraphics[scale=.52]{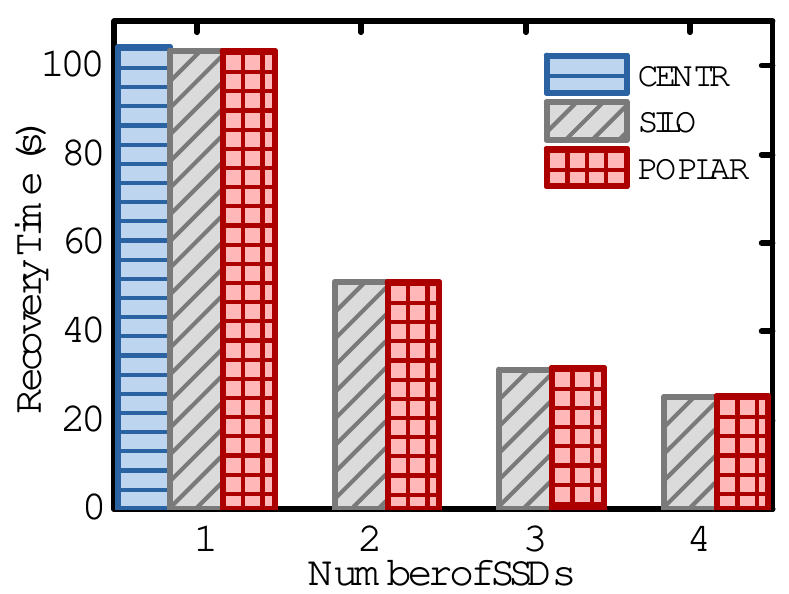}}
	\subfigure[TPC-C\label{fig:recovery-scal-tpcc}]{\includegraphics[scale=.52]{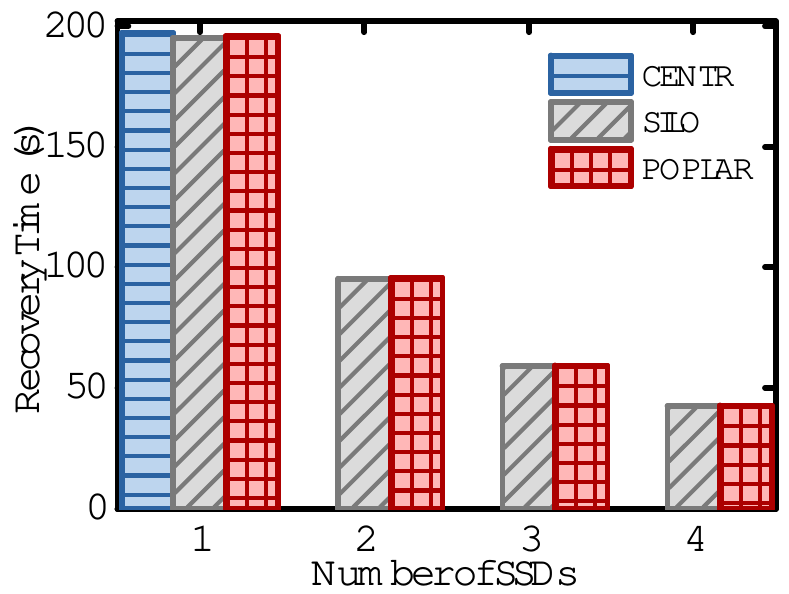}}
	\caption{Scalability of recovery. \para and \silo improve the recovery performance by leveraging parallelism of multiple SSDs. } 
	\label{fig::recovery-scal}  
\end{figure}

\textbf{Recovery scalability.}
To investigate the scalability of the recovery algorithm in different logging approaches, 
we measure the total recovery time by varying the number of SSDs. 
For both YCSB and TPC-C, each experiment must recovery the same checkpoints and log files 
that evenly distributed across the used I/O devices. 
The results are presented in Figure~\ref{fig::recovery-scal}. 
In all workloads, when only a single SSD exists, 
the recovery performance of \para and \silo is almost the same as that of \centralized. 
As the number of SSDs increases, \para and \silo significantly speed up the recovery process by leveraging parallelism of the multiple I/O devices. 
In the graph, the recovery time of approaches is proportional to the number of SSDs, except for \centralized. 
Hence, \para and \silo have better scalability than \centralized.

%% file: relatedwork.tex
\section{Related Work}\label{sec:relatedwork}
\textbf{Centralized logging.} 
ARIES-Logging~\cite{VLDB:mohan1992aries} is widely used in many traditional 
database systems such as DB2, MySQL and PostgreSQL, 
but its centralized design seriously hurts performance of transactional database systems. 
Johnson et al.~\cite{VLDB:johnson2010aether} adequately analyze bottlenecks of ARIES 
logging in the setting with multi-core processors, 
and indicate the contention on the centralized log buffer is the  most important
issue to scalability.  
To eliminate the serialization bottleneck,  Aether~\cite{VLDB:johnson2012scalability} designs a scalable log 
buffer, which aggregates requests from multiple worker threads and supports parallel log insertion. 
Hekaton~\cite{VLDB:diaconu2013hekaton}, ERMIA~\cite{VLDB:kim2016ermia} and Deuteronomy~\cite{VLDB:levandoski2015high} use an atomic instruction instead of a global lock to calculate \lsn for each log record. 
And \textsc{Eleda}~\cite{VLDB:jung2017scalable} proposes a highly concurrent data structure to track \lsn holes caused by concurrent log insertion. 

In addition, database systems must ensure that a transaction's log records have been durable before the transaction commits. 
Flushing log records to a slower storage device damages the throughput and commit latency of transactions. 
Group commit~\cite{VLDB:hagmann1987reimplementing} reduces pressure on storage device by aggregating multiple log records into a single IO operation. 
Asynchronous commit~\cite{VLDB:asynchronouscommit} allows transactions to be committed without waiting for their log records to be persisted. 
This method significantly promotes performance but at the cost of sacrificing the durability. 
Early lock release (ELR)~\cite{VLDB:dewitt1984implementation} allows that a transaction releases locks before persisting its log records, which reduces the lock contention cased by log flushing. 

As system load increases, database systems need to force a large number of log records to a single storage device,  
so that the limited bandwidth becomes the foremost impediment to performance. 
To address this problem, 
H-Store~\cite{VLDB:kallman2008h} employs command 
logging~\cite{VLDB:malviya2014rethink}, 
which only records a procedure identifier and query parameters for a 
transaction. 
The logical log records incur low overhead for logging, 
but significantly slow down the recovery process as crash recovery manager 
needs to re-execute all transactions in a serial order. 

\textbf{Parallel logging.} 
There are many approaches which enable parallel logging to eliminate
the bottlenecks of centralized logging. 
Silo~\cite{VLDB:tu2013speedy,VLDB:zheng2014fast} presents a fast logging and 
recovery algorithm by using multiple log buffers and storage devices. 
To avoid centralized contention and ensure the correctness of recovery, Silo adopts a group 
commit mechanism based on coarse-grained epoch. 
As the emergence of non-volatile memory (NVM), 
there are many NVM-based logging used to improve logging performance. 
PCMLogging~\cite{VLDB:gao2011pcmlogging}, SCM-based Logging~\cite{VLDB:fang2011high} and NV-Logging~\cite{VLDB:huang2014nvram} establish log buffer in NVM and allow worker threads to write log records into NVM in parallel. 
Wang et al.~\cite{VLDB:wang2014scalable} propose a distributed logging, 
which uses multiple log buffers on NVM to persist log records. 
To ensure correctness of recovery, 
it uses a global sequence number (\gsn) to track all dependencies between
transactions, 
and uses a passive group commit to protect transaction committing. 
All works manage to parallelize the transaction logging but is not suitable for slower storage devices due to frequent log flushing.

%% file: conclusion.tex
\section{Conclusion}\label{sec:conclusion}
In this work, we proposed a new transaction logging \myarch based on our defined recoverability for a crash recovery manager. 
\myarch breaks the strongly sequential constraints in centralized logging and allows log records to be written into multiple storage devices in parallel. 
To guarantee the correctness of recovery, 
it uses a scalable sequence number (\mylsn) to track WAW and RAW dependencies 
between transactions 
and allows transactions to be committed after its log record, along with log records of its RAW  predecessors, have been durable. 
After a system crash, \myarch can concurrently recover log records in \mylsn order to restore a consistent state.  
Our evaluation results demonstrate that \myarch can achieve high scalability and good performance in terms of throughput and commit latency.

%% file: ms.bbl
\begin{thebibliography}{10}

\bibitem{VLDB:agigarnvdimm}
Agigatech. nagigaram non-volatile dimms.
\newblock \url{http://www.agigatech.com/products/agigaram-nvdimms/}.

\bibitem{VLDB:DBx1000}
Dbx1000.
\newblock \url{https://github.com//yxymit/DBx1000}.

\bibitem{VLDB:asynchronouscommit}
Postgresql. asynchronous commit.
\newblock \url{https://www.postgresql.org/docs/8.3/wal-async-commit.html/}.

\bibitem{VLDB:vikingnvdimm}
Viking technology. ddr4 nvdimm.
\newblock \url{http://www.vikingtechnology.com/products/nvdimm/ddr4-nvdimm/}.

\bibitem{TPCC}
Tpc benchmark c standard specification (revision 5.11).
\newblock Transaction Processing Performance Council., Februay, 2010.

\bibitem{VLDB:bernstein1987concurrency}
P.~A. Bernstein, V.~Hadzilacos, and N.~Goodman.
\newblock {\em Concurrency Control and Recovery in Database Systems}.
\newblock Addison-Wesley, 1987.

\bibitem{VLDB:cooper2010benchmarking}
B.~F. Cooper, A.~Silberstein, E.~Tam, et~al.
\newblock Benchmarking cloud serving systems with {YCSB}.
\newblock In {\em SoCC}, pages 143--154, 2010.

\bibitem{VLDB:dewitt1984implementation}
D.~J. DeWitt, R.~H. Katz, F.~Olken, L.~D. Shapiro, M.~Stonebraker, and D.~A.
  Wood.
\newblock Implementation techniques for main memory database systems.
\newblock In {\em SIGMOD}, pages 1--8, 1984.

\bibitem{VLDB:diaconu2013hekaton}
C.~Diaconu, C.~Freedman, E.~Ismert, et~al.
\newblock Hekaton: Sql server's memory-optimized oltp engine.
\newblock In {\em SIGMOD}, pages 1243--1254. ACM, 2013.

\bibitem{VLDB:fang2011high}
R.~Fang, H.~Hsiao, B.~He, C.~Mohan, and Y.~Wang.
\newblock High performance database logging using storage class memory.
\newblock In {\em ICDE}, pages 1221--1231, 2011.

\bibitem{VLDB:gao2011pcmlogging}
S.~Gao, J.~Xu, B.~He, B.~Choi, and H.~Hu.
\newblock Pcmlogging: reducing transaction logging overhead with pcm.
\newblock In {\em CIKM}, pages 2401--2404. ACM, 2011.

\bibitem{VLDB:hagmann1987reimplementing}
R.~B. Hagmann.
\newblock Reimplementing the cedar file system using logging and group commit.
\newblock In {\em SOSP}, pages 155--162, 1987.

\bibitem{VLDB:huang2014nvram}
J.~Huang, K.~Schwan, and M.~K. Qureshi.
\newblock Nvram-aware logging in transaction systems.
\newblock {\em VLDB}, 8(4):389--400, 2014.

\bibitem{VLDB:johnson2010aether}
R.~Johnson, I.~Pandis, R.~Stoica, et~al.
\newblock Aether: a scalable approach to logging.
\newblock {\em VLDB}, 3(1-2):681--692, 2010.

\bibitem{VLDB:johnson2012scalability}
R.~Johnson, I.~Pandis, R.~Stoica, et~al.
\newblock Scalability of write-ahead logging on multicore and multisocket
  hardware.
\newblock {\em PVLDB}, 21(2):239--263, 2012.

\bibitem{VLDB:jung2017scalable}
H.~Jung, H.~Han, and S.~Kang.
\newblock Scalable database logging for multicores.
\newblock {\em VLDB}, 11(2):135--148, 2017.

\bibitem{VLDB:kallman2008h}
R.~Kallman, H.~Kimura, J.~Natkins, A.~Pavlo, et~al.
\newblock H-store: a high-performance, distributed main memory transaction
  processing system.
\newblock {\em VLDB}, 1(2):1496--1499, 2008.

\bibitem{VLDB:kim2016ermia}
K.~Kim, T.~Wang, R.~Johnson, and I.~Pandis.
\newblock Ermia: Fast memory-optimized database system for heterogeneous
  workloads.
\newblock In {\em Proceedings of the 2016 International Conference on
  Management of Data}, pages 1675--1687. ACM, 2016.

\bibitem{VLDB:levandoski2015high}
J.~Levandoski, D.~Lomet, S.~Sengupta, R.~Stutsman, and R.~Wang.
\newblock High performance transactions in deuteronomy.
\newblock {\em CIDR}, 2015.

\bibitem{VLDB:li1993post}
X.~Li and M.~H. Eich.
\newblock Post-crash log processing for fuzzy checkpointing main memory
  databases.
\newblock In {\em ICDE}, pages 117--124. IEEE, 1993.

\bibitem{VLDB:malviya2014rethink}
N.~Malviya, A.~Weisberg, S.~Madden, and M.~Stonebraker.
\newblock Rethinking main memory oltp recovery.
\newblock In {\em ICDE}, pages 604--615. IEEE, 2014.

\bibitem{VLDB:mohan1992aries}
C.~Mohan, D.~Haderle, B.~Lindsay, et~al.
\newblock Aries: a transaction recovery method supporting fine-granularity
  locking and partial rollbacks using write-ahead logging.
\newblock {\em TODS}, 17(1):94--162, 1992.

\bibitem{VLDB:thomas1979lww}
R.~Thomas.
\newblock A majority consensus approach to concurrency control for multiple
  copy databases.
\newblock {\em ACM Trans. Database Syst.}, 4(2):180--209, 1979.

\bibitem{VLDB:tu2013speedy}
S.~Tu, W.~Zheng, E.~Kohler, B.~Liskov, and S.~Madden.
\newblock Speedy transactions in multicore in-memory databases.
\newblock In {\em SOSP}, pages 18--32. ACM, 2013.

\bibitem{VLDB:wang2014scalable}
T.~Wang and R.~Johnson.
\newblock Scalable logging through emerging non-volatile memory.
\newblock {\em VLDB}, 7(10):865--876, 2014.

\bibitem{VLDB:yu2016tictoc}
X.~Yu, A.~Pavlo, D.~Sanchez, and S.~Devadas.
\newblock Tictoc: Time traveling optimistic concurrency control.
\newblock In {\em SIGMOD}, pages 1629--1642. ACM, 2016.

\bibitem{VLDB:zhang2015study}
Y.~Zhang and S.~Swanson.
\newblock A study of application performance with non-volatile main memory.
\newblock In {\em MSST}, pages 1--10. IEEE, 2015.

\bibitem{VLDB:zheng2014fast}
W.~Zheng, S.~Tu, E.~Kohler, and B.~Liskov.
\newblock Fast databases with fast durability and recovery through multicore
  parallelism.
\newblock In {\em OSDI}, volume~14, pages 465--477, 2014.

\end{thebibliography}
